\documentclass[aps,pre,twocolumn,showpacs,longbibliography]{revtex4-1}
\usepackage{bbding}
\usepackage{amsmath}
\usepackage{amsthm}
\usepackage{graphicx}
\usepackage{subfigure}
\usepackage{amssymb}
\usepackage{graphicx}
\usepackage{dcolumn}
\usepackage{bm}
\usepackage{float}
\usepackage[colorlinks,linkcolor=blue,citecolor=blue,urlcolor=blue,hyperindex,breaklinks]{hyperref}

\begin{document}

\title{Quantum thermal transistor based on the
qubit-qutrit coupling }
\author{Bao-qing Guo}
\author{Tong Liu}
\author{Chang-shui Yu}
 \email{ycs@dlut.edu.cn}
 \affiliation{School of Physics, Dalian University of Technology, Dalian 116024, China}%
\date{\today }

\begin{abstract}
A quantum thermal transistor is designed by the strong coupling between one qubit and one qutrit which are in contact with three heat baths with different temperatures. The thermal behavior is analyzed based on the master equation by both the numerical and the approximately analytic methods. It is shown that the thermal transistor, as a three-terminal device, allows a weak \textit{modulation} heat current (at the modulation terminal) to \textit{switch on/off} and effectively modulate the heat current between the other two terminals. In particular, the weak \textit{modulation} heat current can induce the strong heat current between the other two terminals with the multiple-region \textit{amplification} of heat current. Furthermore, the heat currents are quite robust to the temperature (current) fluctuation at the lower-temperature terminal within certain range of temperature, so it can behave as a heat current \textit{stabilizer}.
\end{abstract}

\pacs{03.65.Ta, 03.67.-a, 05.30.-d, 05.70.-a}
\maketitle

\affiliation{School of Physics and Optoelectronic Technology, Dalian University of
Technology, Dalian 116024, China}

\section{Introduction}

The diode \cite{ioablbttpm} and the transistor \cite{658753} which directly led to the revolution of the electronic information in the last century are the important components that realize the management of the electronic transport. Diodes are two terminal electronic devices that guide the
electric conduction based on the direction of the electric current, and the transistors with three terminals utilize the electric current at one
terminal to control the electric conduction between the other two terminals to realize three basic functions: a switch, an amplifier, or a modulator. In a similar manner, the thermal devices were expected to be developed for the potential management of the heat currents. It was shown in experiment that the heat currents could be switched on/off in various materials such as the carbon nanotube structures \cite{Chang2006Solid} and so on \cite%
{ref:njp.10.083016,doi:10.1063/1.3253712,van2012Tuning,doi:10.1063/1.3697673}, and the similar functions for heat as the diode or the transistor were shown by the VO2 \cite%
{doi:10.1063/1.4941405,doi:10.1063/1.4829618,doi:10.1063/1.4825168,doi:10.1063/1.4905132,Benabdallah2014Near,doi:10.1063/1.4916730}.

With the increasing interests in the quantum thermodynamics, it paves the way for studying the macroscopic thermodynamic laws at the quantum level and
designing the thermal machines/devices in the quantum systems. For examples, the Fourier laws for the heat conduction and the second thermodynamic law
were studied in the various systems \cite%
{ref:vofliodmclwaii,ref:flfacocaouecn,ref:boflintc,ref:qteafl,ref:hciaodas,ref:rociodhc,ref:ahctawl,ref:tlatqtmvtslot,ref:fot,ref:qrattlot,ref:ctpomdai}
and the quantum heat engine and refrigerator \cite%
{ref:podheahpift,ref:qtcc,ref:ccfao,ref:mhp,ref:olsdcear,ref:aqmheoift,ref:tqheahp, ref:opordqr,ref:cqoc,ref:heiftgbme,ref:qfshetoiamwif,ref:qtcaqhe,ref:hsctmbtspr,ref:retscqritscr,ref:PhysRevE.96.012122,ref:sqarwrc,ref:siheamp,ref:asahe,ref:tlmahe,ref:tqosaamothe,ref:tsrcrme} have also been designed. In particular, it is shown that not only the heat
logic gates \cite{ref:tlgcwp}, the thermal memory \cite{ref:tmasopi}, the thermal ratchet \cite{ref:otr,ref:mwaaqhr}, and thermometer \cite%
{ref:PhysRevLett.119.090603}, but also the analogues of the electronic devices, the thermal rectifier \cite%
{ref:oritucr,ref:tritnqds,ref:tdrohf,ref:scftriggm,ref:trigm,doi:10.1063/1.3253712, ref:fqtoald,ref:fritqxxzc,ref:chfatracttls,ref:pttar}, the transistor \cite%
{Benabdallah2014Near,doi:10.1063/1.4916730,ref:pttar,ref:qtt,ref:tthfsam,Li2006Negative,Komatsu2011Thermal} have been theoretically proposed and investigated extensively. It is worth emphasizing that the thermal devices with only several levels have also been proposed such as a thermal rectifier made of only one quantum dot with high in-plane magnetic fields \cite{ref:njp.10.083016}, optimal rectification
consisting of two two-level systems (TLSs) in a magnetic field \cite%
{ref:oritucr}, a quantum thermal transistor with three TLSs \cite%
{ref:qtt,ref:zna.72.163} and so on. Recently artificial atoms such as superconducting circuits and spins in solids \cite{ref:you2011atomic,ref:naturalandartificialatoms} provide a novel and flexible method to investigate quantum thermodynamics or to design quantum thermal machines\cite{ref:PhysRevB.94.235420,ref:coupledqubitsasaswitch}.  In this sense, how to design the thermal device in the small system and how to improve the various performance indices of some particular functions become the significant topics. 

In this paper, we design the thermal transistor by employing the only strong qubit-qutrit coupling. Our thermal transistor consists of one qubit and one qutrit which interact with three heat baths with different temperatures. The master equation governing the dynamic evolution of the open system is
derived and solved numerically and approximately analytically. It is shown that our thermal transistor allows a weak \textit{modulation} heat current to switch on/off and effectively modulate the heat current between the other two terminals. In particular,  the weak \textit{modulation} heat current can induce the strong heat current between the other two terminals, which realizes the typical function of a transistor---the amplification. Moreover, it is shown that the heat currents are quite robust to the temperature change at the lower-temperature terminal within certain range of temperature, so it can be used to realize the heat current stabilization subject to the temperature fluctuation at the lower-temperature terminal. The distinct features of our transistor are (1) at the \textit{off} state, the \textit{modulation} heat current has a large allowable region and a quite weak heat current; (2) the transistor has multiple (stable or sensitive) amplification regions which have different (very large) amplification factors; (3) it is robust to the temperature fluctuation at the lower-temperature terminal. The remaining of this paper is organized as follows. In Sec.~\ref{sec:modelandME}, we derive the master equation that governs that evolution of our proposed open system. In Sec.~\ref{sec:steadystateandcurrent}, we solve the master equation and calculate the heat currents at the steady state. In Sec.~\ref%
{sec:transistoreffectsanddiscussions}, we analyze the thermal behavior and show how our system behaves as a quantum thermal transistor. Finally, some discussions and the conclusion are given in Sec.~\ref{sec:conclusion}. 
\section{\label{sec:modelandME} The model and the dynamics}

\begin{figure}[tbp]
\centering
\includegraphics[width=0.75\columnwidth]{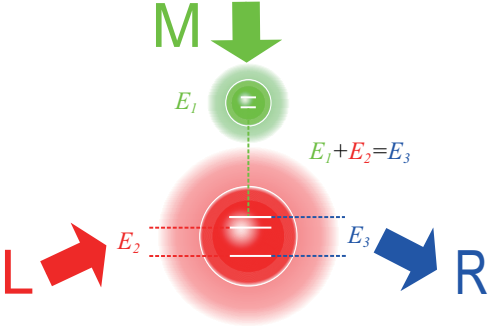}
\caption{(Colour online) A \textit{modulation} qubit in contact with the heat  bath \textit{M} and a qutrit connected to a heat bath \textit{L} and the heat bath \textit{R}  are strongly coupled with each other, which acts as a quantum thermal device. The ground state and the first excited state of the qutrit are greatly separated, while the transition between the first and the second excited states is resonantly coupled to the \textit{modulation} qubit.}
\label{fig:modeldiagram}
\end{figure}

Our model as sketched in Fig.~\ref{fig:modeldiagram}, consists of a qubit as the \textit{modulation} qubit which is in contact with a heat bath \textit{M}
with the temperature $T_M$ and simultaneously interacts with a \textit{target%
} qutrit which is in contact with a heat bath \textit{L} and a heat bath 
\textit{R} with their temperatures denoted by $T_L$ and $T_R$. In the
following, we will demonstrate that the weak heat current through the 
\textit{modulation} qubit can be used to switch on/off, modulate and
stabilize the heat currents through the target qutrit from the \textit{L}
bath to the \textit{R} bath (or in the opposite direction). In particular,
as a key feature of the transistor, it can be seen that the
amplification of the weak heat current can also be realized by this model.

To show this, let's turn to the dynamical procedure of our model. For
simplicity, we would like to suppose that the ground-state energies are zero
for both the qubit and the qutrit. Let ${\left\vert 1\right\rangle }_{1}={%
\left[ 1,0\right] }^{T}$ denote the excited state of the qubit with the
energy $E_{1}$ and ${\left\vert 1\right\rangle }_{2}={\left[ 0,1,0\right] }%
^{T}$ and ${\left\vert 2\right\rangle }_{2}={\left[ 1,0,0\right] }^{T}$
denote two excited states of the qutrit, respectively, corresponding to the
energies $E_{2}$ and $E_{3}$. In addition, we suppose the qubit and the
qutrit interact with each other via the Hamiltonian 
\begin{equation}
H_{I}=g(\left\vert 11\right\rangle \left\langle 02\right\vert +\left\vert
02\right\rangle \left\langle 11\right\vert )
\end{equation}%
with $g$ representing the coupling strength, so the Hamiltonian of the
bipartite interacting system reads 
\begin{equation}
H_{S}=H_{0}+H_{I},
\end{equation}%
where the free Hamiltonian is 
\begin{equation}
H_{0}=E_{1}{\left\vert 1\right\rangle }_{1}\left\langle 1\right\vert +E_{2}{%
\left\vert 1\right\rangle }_{2}\left\langle 1\right\vert +E_{3}{\left\vert
2\right\rangle }_{2}\left\langle 2\right\vert .
\end{equation}%
Here we consider the resonant coupling, i. e., $E_{1}+E_{2}=E_{3}$ and we
set the Boltzmann constant and the Plank constant to be unit, i. e., $\hbar
=k_{B}=1$. Now let's consider the qubit-qutrit system interacts with three
heat baths which are described by the quantized radiation field. The free
Hamiltonian of the three baths reads 
\begin{equation}
H_{\mu }=\sum_{k}\omega _{\mu k}b_{\mu k}^{\dagger }b_{\mu k},\mu =L,M,R,
\end{equation}%
where $\omega _{\mu k}$ and $b_{\mu k}$ denote the frequency and the
annihilation operator of the bath modes with $[b_{\mu k}^{\dagger },b_{\nu
k^{^{\prime }}}]=\delta _{\mu ,\nu }\delta _{k,k^{^{\prime }}},[b_{\mu
k}^{\dagger },b_{\nu k^{^{\prime }}}^{\dagger }]=0,[b_{\mu k},b_{\nu
k^{^{\prime }}}]=0$. The interaction Hamiltonian between the system and the
baths is given by 
\begin{align}
H_{SB}=& \sum_{k}f_{Lk}(b_{Lk}^{\dagger }{\left\vert 0\right\rangle }%
_{2}\left\langle 1\right\vert +b_{Lk}{\left\vert 1\right\rangle }%
_{2}\left\langle 0\right\vert ) \\
+& \sum_{k}f_{Mk}(b_{Mk}^{\dagger }{\left\vert 0\right\rangle }%
_{1}\left\langle 1\right\vert +b_{Mk}{\left\vert 1\right\rangle }%
_{1}\left\langle 0\right\vert )  \notag \\
+& \sum_{k}f_{Rk}(b_{Rk}^{\dagger }{\left\vert 0\right\rangle }%
_{2}\left\langle 2\right\vert +b_{Rk}{\left\vert 2\right\rangle }%
_{2}\left\langle 0\right\vert ),  \notag
\end{align}%
where $f_{\mu k}$ denotes the coupling constants between the $k$th mode in
the $\mu $th bath and the corresponding energy levels of the system. Thus
the Hamiltonian of the whole open system can be given as 
\begin{equation}
H_{total}=H_{S}+\sum_{\mu }H_{\mu }+H_{SB}.  \label{Hami}
\end{equation}

Based on such a Hamiltonian (\ref{Hami}), one can derive the dynamical
equation of the open system, i. e., the master equation \cite{ref:ttooqs}.
One can note that $H_{S}$ can be diagonalized as $H_{S}=\sum_{i=1,2,...,6}{%
\lambda }_{i}\left\vert {\lambda }_{i}\right\rangle \left\langle {\lambda }%
_{i}\right\vert $, where the eigenvalues are given by ${\left\vert \lambda
\right\rangle }^{T}=[\lambda _{1},\lambda _{2},...,\lambda
_{6}]=[E_{1}+E_{3},E_{3}-g,E_{1},E_{3}+g,E_{2},0]$, and the corresponding
eigenstates are 
\begin{equation}
\begin{array}{ccc}
\left\vert \lambda _{1}\right\rangle =\left\vert 12\right\rangle , \ 
\left\vert \lambda _{2}\right\rangle =\frac{1}{\sqrt{2}}(\left\vert
11\right\rangle -\left\vert 02\right\rangle ), \ 
\left\vert \lambda _{3}\right\rangle =\left\vert 10\right\rangle ,\\  
\left\vert \lambda _{4}\right\rangle =\frac{1}{\sqrt{2}}(\left\vert
11\right\rangle +\left\vert 02\right\rangle ), \ 
\left\vert \lambda _{5}\right\rangle =\left\vert 01\right\rangle , \ 
\left\vert \lambda _{6}\right\rangle =\left\vert 00\right\rangle .%
\end{array}%
\end{equation}%
In the $H_{S}$ presentation, the interaction Hamiltonian $H_{SB}$ can be
rewritten as 
\begin{equation*}
H_{SB}=\sum_{\mu ,k,j}f_{\mu k}(b_{\mu k}^{\dagger }V_{\mu l}(\omega _{\mu
l})+b_{\mu k}V_{\mu l}^{\dagger }(\omega _{\mu l})),
\end{equation*}%
where $V_{\mu l}(\omega _{\mu l})$ stands for the eigenoperator of $H_{S}$
corresponding to the eigenfrequency $\omega _{\mu l}$ with the relation $%
[H_{S},V_{\mu l}(\omega _{\mu l})]=-\omega _{\mu l}V_{\mu l}(\omega _{\mu
l}) $. The concrete expressions of the eigenoperators are given in Appendix~%
\ref{appx:sec:eigenoperators}. It is clear that the transitions $\left\vert
\lambda _{3}\right\rangle \leftrightarrow \left\vert \lambda
_{2}\right\rangle $, $\left\vert \lambda _{6}\right\rangle \leftrightarrow
\left\vert \lambda _{5}\right\rangle $, $\left\vert \lambda
_{3}\right\rangle \leftrightarrow \left\vert \lambda _{4}\right\rangle $ are
driven by the bath $L$, $\left\vert \lambda _{6}\right\rangle
\leftrightarrow \left\vert \lambda _{3}\right\rangle $, $\left\vert \lambda
_{5}\right\rangle \leftrightarrow \left\vert \lambda _{2}\right\rangle $, $%
\left\vert \lambda _{4}\right\rangle \leftrightarrow \left\vert \lambda
_{1}\right\rangle $, $\left\vert \lambda _{5}\right\rangle \leftrightarrow
\left\vert \lambda _{4}\right\rangle $, $\left\vert \lambda
_{2}\right\rangle \leftrightarrow \left\vert \lambda _{1}\right\rangle $ are
driven by the bath $M$, and $\left\vert \lambda _{6}\right\rangle
\leftrightarrow \left\vert \lambda _{2}\right\rangle $, $\left\vert \lambda
_{6}\right\rangle \leftrightarrow \left\vert \lambda _{4}\right\rangle $, $%
\left\vert \lambda _{3}\right\rangle \leftrightarrow \left\vert \lambda
_{1}\right\rangle $ are driven by the bath $R$. Following the standard
procedure \cite{ref:ttooqs}, within the Born-Markovian approximation and the
secular approximation, one can obtain the master equation in Schr\"{o}dinger
picture as 
\begin{equation}
\dot{\rho}=-{i}[H_{S},\rho ]+\mathcal{L}_{L}[\rho ]+\mathcal{L}_{M}[\rho ]+%
\mathcal{L}_{R}[\rho ],  \label{mas1}
\end{equation}%
where the dissipator $\mathcal{L}_{\mu }[\rho ]$ is given by 
\begin{gather}
\mathcal{L}_{\mu }[\rho ]=\sum_{l}J_{\mu }(-\omega _{\mu
l})[2V_{\mu l}(\omega _{\mu l})\rho V_{\mu l}^{\dagger }(\omega _{\mu l}) 
\notag \\
-\{V_{\mu l}^{\dagger }(\omega _{\mu l})V_{\mu l}(\omega _{\mu l}),\rho
\}]+J_{\mu }(+\omega _{\mu l})[2V_{\mu l}^{\dagger }(\omega _{\mu l})\rho
V_{\mu l}(\omega _{\mu l})  \notag \\
-\{V_{\mu l}(\omega _{\mu l})V_{\mu l}^{\dagger }(\omega _{\mu l}),\rho \}],
\end{gather}%
with the spectral densities defined by 
\begin{align}
& J_{\mu }(+\omega _{\mu l})=\gamma _{\mu }(\omega _{\mu l})n(\omega _{\mu
l}), \\
& J_{\mu }(-\omega _{\mu l})=\gamma _{\mu }(\omega _{\mu l})[n(\omega _{\mu
l})+1],
\end{align}%
and the average photon number given by 
\begin{equation}
n(\omega _{\mu l})=\frac{1}{e^{\frac{\omega _{\mu l}}{T_{\mu }}}-1}
\end{equation}%
corresponding to the frequency $\omega _{\mu l}$ and the temperature $T_{\mu
}$. Due to the secular approximation, it requires $\gamma _{\mu }(\omega
_{\mu l})\ll \{|\omega _{\mu l}-\omega _{\mu l^{^{\prime }}}\pm 2g|,g\}$
which implies the strong internal coupling. In addition, we assume $\gamma
_{\mu }(\omega _{\mu l})=\gamma _{\mu }$ independent of the transition
frequency for simplicity.

\section{ \label{sec:steadystateandcurrent} Steady state of the open system
and the heat currents}

\begin{figure}[tbp]
\centering
\includegraphics[width=0.75\columnwidth]{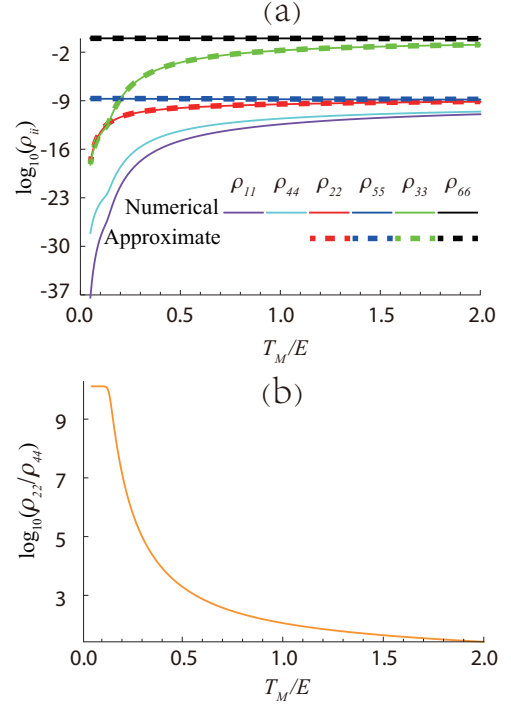}
\caption{(Colour online) (a) Numerical (solid lines) and approximate (dashed lines) populations versus $T_{M}$ for the steady state. The lines from top to bottom at $T_{M}/E=2$ correspond to $\rho_{66}$, $\rho_{33}$, $\rho_{55}$, $\rho_{22}$, $\rho_{44}$, $\rho_{11}$. (b) The ratio $\rho_{22}/\rho_{44}$ versus $T_{M}$ for the steady state.  Here $E_{1}=4E$, $E_{2}=40E$%
, $E_{3}=E_{1}+E_{2}=44E$, $g=0.75 E_{1}$, $\protect\gamma =0.01 E_{1}$, $T_{L}=2E$, $T_{R}=0.2E$. For both figures, one can find
 that  $\rho_{22}$  is less than $\rho_{33}$, $\rho_{55}$ and $\rho_{66}$, in the meanwhile, $\rho_{11}$ is less than $\rho_{44}$ which is much less than $\rho_{22}$. In this sense, $\rho_{11}$  and $\rho_{44}$ can be safely neglected to a good approximation.}
\label{fig:populationsTM}
\end{figure}

To demonstrate the functions of a thermal transistor, we need to study the
steady-state thermal behavior of the open system. So the first key task is
to find the steady solution of the master equation Eq.~(\ref{mas1}), namely,
to solve $\dot{\rho}^{S}=0$ (or Eq.~(\ref{mas1}) with $t\rightarrow \infty $).
To do so, we rewrite the master equation for the steady state as 
\begin{gather}
\sum_{\mu =M,L,R}\mathbf{M}_{\mu } \left\vert \rho \right\rangle =0,  \label{eq:rhodiagonal} \\
\rho _{ij}^{S} =0,i\neq j,  \notag
\end{gather}%
where $\left\vert \rho \right\rangle^{T} =[\rho _{11}^{S},\rho _{22}^{S},\rho
_{33}^{S},\rho _{44}^{S},\rho _{55}^{S},\rho _{66}^{S}]$
with {\allowdisplaybreaks[4]
\begin{align}
\mathbf{M}_{L}& =C_{2,1;3,2}\mathbf{J}_{L1}C_{2,1;3,2}^{\dagger }  \notag \\
& +2C_{5,1;6,2}\mathbf{J}_{L2}C_{5,1;6,2}^{\dagger }+C_{3,1;4,2}\mathbf{J}%
_{L3}C_{3,1;4,2}^{\dagger }, \\
\mathbf{M}_{M}& =2C_{3,1;6,2}\mathbf{J}_{M1}C_{3,1;6,2}^{\dagger }  \notag \\
& +C_{1,1;4,2}\mathbf{J}_{M2}C_{1,1;4,2}^{\dagger }+C_{2,1;5,2}\mathbf{J}%
_{M2}C_{2,1;5,2}^{\dagger }  \notag \\
& +C_{1,1;2,2}\mathbf{J}_{M3}C_{1,1;2,2}^{\dagger }+C_{4,1;5,2}\mathbf{J}%
_{M3}C_{4,1;5,2}^{\dagger }, \\
\mathbf{M}_{R}& =C_{2,1;6,2}\mathbf{J}_{R1}C_{2,1;6,2}^{\dagger }  \notag \\
& +C_{4,1;6,2}\mathbf{J}_{R2}C_{4,1;6,2}^{\dagger }+C_{1,1;3,2}\mathbf{J}%
_{R3}C_{1,1;3,2}^{\dagger }.
\end{align}%
}Here $\mathbf{J}_{\mu l}=\left\vert 2\right\rangle \left\langle
2\right\vert \otimes 
\begin{pmatrix}
-A_{\mu l} & B_{\mu l} \\ 
A_{\mu l} & -B_{\mu l}%
\end{pmatrix}%
$ with $A_{\mu l}=\gamma _{\mu }(n(\omega _{\mu l})+1)$ and $B_{\mu
l}=\gamma _{\mu }n(\omega _{\mu l})$, and $C_{i,j;m,n}=\left\vert
i\right\rangle \left\langle j\right\vert +\left\vert m\right\rangle
\left\langle n\right\vert $ with $\left\{ \left\vert i\right\rangle \right\} 
$ representing the orthonormal basis of $6$-dimensional Hilbert space. One can find that Eq.~(\ref{eq:rhodiagonal}) is analytically solvable, but the concrete expression is so tedious that it is impossible to present explicitly here. So we make some reasonable approximations in order to give an explicit presentation. Here all the involved parameters are taken as $E_{1}=4E$, $E_{2}=40E$, $E_{3}=E_{1}+E_{2}=44E$, $g=0.75E_{1}$, $\gamma =0.01E_{1}$, $T_{L}=2E$, $T_{R}=0.2E$ and $T_{M}<$ $%
T_{L}$. Under this condition, the two higher energy levels $\left\vert \lambda_1\right\rangle$ and $\left\vert \lambda_4\right\rangle$ of $H_{S}$ are difficult
to excite, so one can easily check that the populations of $\rho _{11}^{S}$
and $\rho _{44}^{S}$ are much less than others, which can be seen from Fig.~\ref{fig:populationsTM} (a) and (b). This means that the contributions of these two energy levels $\left\vert \lambda_1\right\rangle$ and $\left\vert \lambda_4\right\rangle$ can be safely neglected to some good approximation. Thus we can replace the irrelevant matrix entries in Eq. (\ref{eq:rhodiagonal}) by \textit{zero}. In this way, the simplified Eq.~(\ref{eq:rhodiagonal}) can be written as 
\begin{gather}
\sum_{\mu =M,L,R}\mathbf{\tilde{M}}_{\mu }\left\vert \tilde{\rho}\right\rangle =0, \\
\rho _{22}^{S}+\rho _{33}^{S}+\rho _{55}^{S}+\rho _{66}^{S}=1,  \notag
\end{gather}%
where $\mathbf{\tilde{M}}_{L} = C_{2,1;3,2}\mathbf{J}_{L1}C_{2,1;3,2}^{\dagger} + 2C_{5,1;6,2}\mathbf{J}_{L2}C_{5,1;6,2}^{\dagger}$, $\mathbf{\tilde{M}}_{M} = 2C_{3,1;6,2}\mathbf{J}_{M1}C_{3,1;6,2}^{\dagger}+C_{1,1;4,2}\mathbf{J}_{M2}C_{1,1;4,2}^{\dagger} +C_{2,1;5,2}\mathbf{J}_{M2}C_{2,1;5,2}^{\dagger}$, $\mathbf{\tilde{M}}_{R} = C_{2,1;6,2}\mathbf{J}_{R1}C_{2,1;6,2}^{\dagger }$ and $\left\vert \tilde{\rho}\right\rangle =[0, \rho _{22}^{S},\rho _{33}^{S}, 0, \rho
_{55}^{S},\rho _{66}^{S}]^{T}$. As a result, one can obtain 
\begin{equation}
\rho _{22}^{S}=\frac{D_{2}}{D},\rho _{33}^{S}=\frac{D_{3}}{D},\rho _{55}^{S}=%
\frac{D_{5}}{D},\rho _{66}^{S}=\frac{D_{6}}{D},  \label{solution}
\end{equation}%
where
{\allowdisplaybreaks[4]
\begin{align}
D_{2}& =2A_{L2}\left[ 2B_{M1}B_{L1}+B_{R1}\left( 2A_{M1}+B_{L1}\right) %
\right] +B_{M2}  \notag \\
& \times\left[ 2B_{M1}B_{L1}+\left( 2A_{M1}+B_{L1}\right) \left(
2B_{L2}+B_{R1}\right) \right] , \\
 D_{3}& =2B_{M1}\left[ 2A_{M2}A_{L2}+A_{R1}(2A_{L2}+B_{M2})\right] +A_{L1}   \notag \\
&\times\left[ 2A_{L2}(2B_{M1}+B_{R1})+B_{M2}(2B_{M1}+2B_{L2}+B_{R1})\right]
, \\
D_{5}& =2B_{L2}\left[ A_{R1}B_{L1}+2A_{M1}(A_{L1}+A_{R1})\right]  +A_{M2}  \notag \\
&\times\left[ 2B_{M1}B_{L1}+(2A_{M1}+B_{L1})(2B_{L2}+B_{R1})\right] , \\
D_{6}& =B_{L1}\left[ 2A_{M2}A_{L2}+A_{R1}(2A_{L2}+B_{M2})\right]+2A_{M1}   \notag \\
& \times\left[ 2A_{M2}A_{L2}+(A_{L1}+A_{R1})(2A_{L2}+B_{M2})\right] , \\
D&=D_{2}+D_{3}+D_{5}+D_{6}.  \notag
\end{align}%
}%
With the solutions of Eq.~(\ref{solution}), one can calculate the heat
currents subject to different baths as \cite%
{ref:PhysRevE.87.012120,ref:PhysRevA.73.052311,ref:qrattlot,ref:PhysRevLett.109.090601}%
\begin{equation}
\dot{Q}_{\mu }={Tr}(H_{S}\mathcal{L}_{\mu }[\rho ^{S}]) \approx \left\langle \lambda
\right\vert \mathbf{\tilde{M}}_{\mu }\left\vert \tilde{ \rho }\right\rangle ,
\label{eq:currentdefinition}
\end{equation}%
which can be explicitly given by 
\begin{align}
\dot{Q}_{L}& =-\omega _{L1}\Gamma _{23}^{L}-2\omega _{L2}\Gamma
_{56}^{L},  \label{eq:currentQL} \\
\dot{Q}_{M}& =-2\omega _{M1}\Gamma _{36}^{M}-\omega _{M2}\Gamma _{25}^{M},
\label{currentQM} \\
\dot{Q}_{R}& =-\omega _{R1}\Gamma _{26}^{R},  \label{eq:currentQR}
\end{align}%
where 
\begin{equation}
\Gamma _{i,j}^{\mu }=\gamma _{\mu }[(n_{\mu }(\lambda _{i}-\lambda
_{j})+1)\rho _{ii}^{S}-n_{\mu }(\lambda _{i}-\lambda _{j})\rho _{jj}^{S}]\label{decayr}
\end{equation}%
denotes the net decay rate from the state $\left\vert i\right\rangle $ to $%
\left\vert j\right\rangle $ due to the coupling with the $\mu $th bath. One
knows that $\dot{Q}_{\mu }>0$ means the heat flows out of the $\mu $th bath
and $\dot{Q}_{\mu }<0$ corresponds to the heat flows into the $\mu $th bath.
It can be easily checked that $\dot{Q}_{L}+\dot{Q}_{M}+\dot{Q}_{R}=0$
corresponding to the energy conservation law. In the next section, we will
show that the heat currents can be effectively controlled and hence our
thermal device can realize the functions of a thermal transistor.

\section{The functions as a transistor \label%
{sec:transistoreffectsanddiscussions}}

Now we will show that the weak \textit{modulation} heat current $\dot{Q}_{M}$
can modulate, switch, and stabilize the output currents through the \textit{%
target} qutrit, moreover, our model can realize the typical function
as a thermal transistor---the amplification of the weak \textit{%
modulation} heat current $\dot{Q}_{M}$, that is, the left current $\dot{Q}%
_{L}$ or the right one $\dot{Q}_{R}$ can become greatly larger than $\dot{Q}%
_{M}$ with a dynamical amplification factor $\alpha $ defined as 
\begin{equation}
\alpha _{L,R}=\frac{\partial \dot{Q}_{L,R}}{\partial \dot{Q}_{M}}.
\label{eq:amplificationfactor}
\end{equation}%
If the amplification factor $\alpha _{L,R}>1$, we can say the transistor
effect is achieved. In particular, the larger $\alpha _{L,R}$ is, the better
transistor effect is obtained. 

\begin{figure}[tbp]
\centering   
\includegraphics[width=0.75%
\columnwidth]{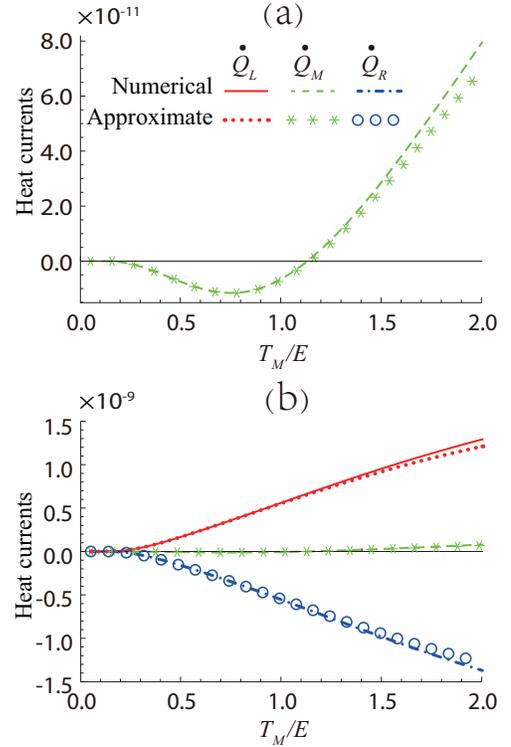}
\caption{(Colour online) Three thermal currents $\dot{Q}_{L}$, $\dot{Q}_{M}$%
, and $\dot{Q}_{R}$ via numerical and approximate methods at steady state
versus $T_{M}$. The parameters are same as in Fig.~\ref{fig:populationsTM}.}
\label{fig:tansistorcurrentsTM}\end{figure}

\textit{Switch}. --In order to show the function as a quantum thermal switch, we plot the three heat currents in Fig.~\ref{fig:tansistorcurrentsTM}. It is obvious that all the three heat currents are very small even close to zero in the low temperature $%
T_M$ regime, i.e., $T_{M}/E \lesssim 0.3$. Especially $\dot{Q}_{L,R}$ in the low temperature $T_M$ regime are much smaller than those in the large $T_M$ regime. Therefore, if the heat currents are neglectfully small, we can think that the heat conduction is prevented between the bath \textit{L} and the bath \textit{R}. In this sense, one can find that our model can be considered to be at the ``\textit{%
off}'' state for $T_{M}/E \lesssim 0.3$. With the increase of $T_M$, the heat currents $\dot{Q}_{L,R}$ are gradually increased, namely, the switch is gradually open and the heat is allowed to transport between the bath \textit{L} and the bath \textit{R}. It is especially noted that if the switch is off, $T_M$ can be taken in a large safe range so long as $T_{M}/E \lesssim 0.3$ is satisfied. Actually, one can always define an exact critical small value  of the allowable heat current based on the practical case. When the heat current value is less than this critical value, one can think the switch is off and when the heat current is larger than the critical value, the switch is on.  

\textit{Modulation}. --The modulation function means the heat current can be controlled continuously from a small value to a large one by another
much smaller continuous current. From Fig.~\ref{fig:tansistorcurrentsTM}, one can find that in the whole range of $T_{M}$, $\dot{Q}_{M}$ always keeps greatly smaller than $\dot{Q}_{L,R}$, while $\dot{Q}_{L,R}$ ranges from a small value (can reach zero) at low $T_{M}$ to a large one for a large $T_{M}$. In this perspective, the two currents $\dot{Q}_{L,R}$ are modulated by a tiny modulation current $\dot{Q}_{M}$ and the modulation function is realized.

\begin{figure}[tbp]
\centering
\includegraphics[width=0.75%
\columnwidth]{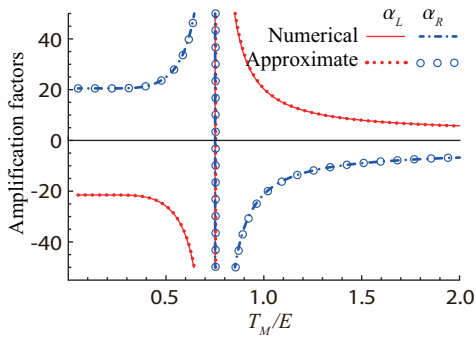}
\caption{(Colour online) Two amplification factors $\protect\alpha_{L}$  and $\protect\alpha_{R}$ via numerical and approximate methods at steady state
versus $T_{M}$. The parameters are same as in Fig.~\protect\ref{fig:populationsTM}.}
\label{fig:factorTM}
\end{figure}

\textit{Amplification}. --The crucial feature of a transistor is
the function of amplification, namely, the weak modulation heat current $\dot{Q}_M$ amplifies (or induces) a strong heat current which transports between the bath  \textit{L} and the bath \textit{R}.  In fact, it is apparent that from Fig.~\ref{fig:tansistorcurrentsTM},  the current $\dot{Q}_M$  varies gently when $0.3 \lesssim T_{M}/E$, but  the currents $\dot{Q}_{L,R}$ are changed  rapidly, which implies the amplification is achieved. However, in order to precisely describe the amplification effect, one has to employ the amplification factor $\alpha$ defined in Eq.~(\ref{eq:amplificationfactor}).  In Fig.~\ref{fig:factorTM}, we present the two amplification factors $\alpha_{L}$ and $\alpha_{R}$ versus $T_{M}$. The two amplification factors are obviously larger than $1$, which shows that the amplification effect indeed exits in our model. At the low
temperature range $0<T_{M}/E<0.5$, the heat current is stably amplified  due to almost the same amplification factors (about $20$). At the range $0.5<T_{M}/E<1$, the amplification factors strongly depend on the temperature $T_{M}$. This can be regarded as a sensitive region which means a tiny change of the modulation current $\dot{Q}_M$ can lead to the drastic change of the currents  $\dot{Q}_{L,R}$. The region $T_M/E>1$ can be considered as the weak stable amplification region. But the factors $\alpha_{L,R}$ are still larger than 1, for example, $\alpha_{L}=5.749$ and $\alpha_{R}=-6.749$ for $T_{M}/E=2.0$. Thus, one can select the proper working region based on what kind of amplification is required  in the practical scenario.

\begin{figure}[tbp]
\centering   
\includegraphics[width=0.75%
\columnwidth]{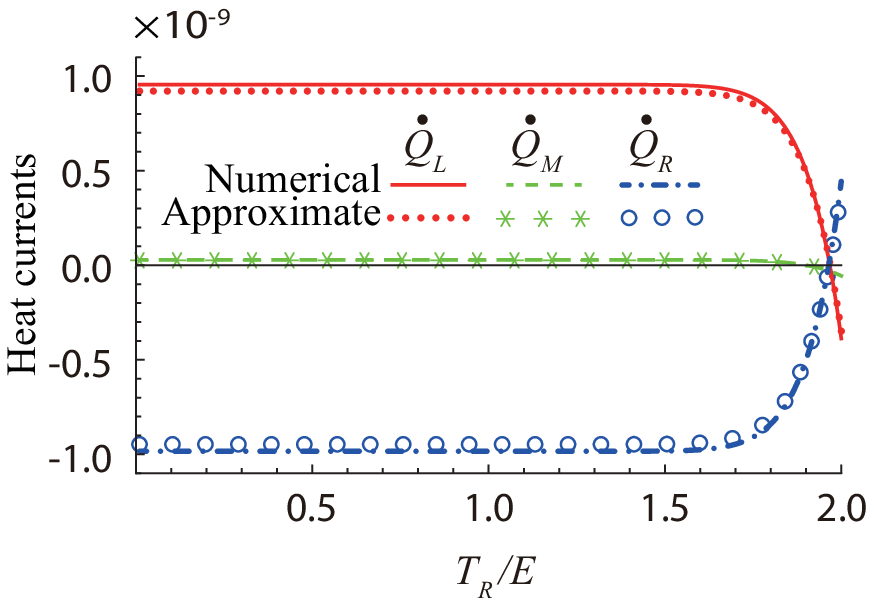}
\caption{(Colour online) Three thermal currents $\dot{Q}_{L}$, $\dot{Q}_{M}$, and $\dot{Q}_{R}$ via numerical and approximate methods at steady state
versus $T_{R}$. Here $T_{M}/E=1.5$ and the parameters are same as in Fig.~\ref{fig:populationsTM}.}
\label{fig:stabilizercurrentsTR}\end{figure}

\textit{Stabilizer}. --In fact, our model can also work for a stabilizer of heat currents, namely,  the heat currents $\dot{Q}_{L,R}$ are not sensitive to the change of the temperature of  $T_R$ (the low-temperature terminal). To illustrate such a function, we plot the three heat currents versus  versus $T_R$ in Fig. \ref{fig:stabilizercurrentsTR}. One can see that when the temperature $T_R$ varies from $0$ to about $T_{R}$ along the horizontal axis, the heat currents $\dot{Q}_{L,R}$ are kept almost in the horizontal lines, that is, there is no obvious change of the heat currents $\dot{Q}_{L,R}$. In fact, a direct understanding of this phenomenon can be obtained by our Eq. (\ref{eq:currentQR}) and Eq. (\ref{decayr}) where the fluctuation of the lower temperature at the $R$ terminal subject to the large transition frequency between the energy levels $\left\vert\lambda_2\right\rangle$ and $\left\vert\lambda_6\right\rangle$ can not lead to the considerable fluctuation of the net decay rate. In other words, the large fluctuations of  $T_R$ cannot drastically influence the heat currents $\dot{Q}_{L,R}$, namely, $\dot{Q}_{L,R}$ in the given temperature range are stabilized. 

 Finally, we emphasize that the thermal transistor proposed in this paper is a thermal device with three terminals. Here we use the heat $Q_M$ as the ``modulation" terminal which controls the heat currents between the other two terminals. What we would like to emphasize is that the choice of the ``modulation" terminal is not unique. In the Appendix~\ref{appx:sec:terminals}, we have numerically studied the cases with $Q_L$ and $Q_R$ as the ``modulation" terminal respectively. It is shown that in both cases our proposed thermal device can realize all the mentioned functions as a transistor including the functions of the switch, the modulation and the amplification. As to the function of the stabilizer, one can also find that if the heat current $Q_L$ as the low-temperature terminal, the heat currents $Q_{L,R}$ can also be stabilized. In fact, throughout of the paper, we intend to fix $T_M$ is a medium temperature between $T_L$ and $T_R$, which is enough for us to show our device as a transistor. If other parameters are selected, one can find that the function as a transistor can be realized in different cases which are not indicated extensively.  We also consider the case of the weak internal coupling. One can find that the transistor effect still exists, but  the price is that the heat currents will be reduced to the very low level (two small). This actually coincides with Refs. \cite{ref:retscqritscr,ref:PhysRevE.96.012122} working in the cooling regime, where the strong internal coupling suppresses the cooling, but the current model works in the heating regime. An intuitive understanding could be that the suppression of cooling implies the enhancement of heating. In addition, one can also find that compared with Ref. \cite{ref:qtt},  we have realized the similar functions with less energy levels and particles with a different mechanism.

\section{Discussion and conclusion \label{sec:conclusion}}
Before the end, we would like to give some discussions about the potential design in the superconducting systems. 
As we know, the superconducting artificial atom provides a possibility to realize a $\Delta$-type system allowing different transitions between the three levels \cite{ref:you2011atomic,ref:naturalandartificialatoms}. One distinct advantage is that the energy gap can be customized freely, for instance, via changing magnetic flux in a circuit QED architecture and  another advantage is  that the coupling between the superconducting artificial atoms can be easily tuned to be strong \cite{ref:niskanen2007quantum,ref:hime2006solid}.  The energy gap of superconducting circuits ranges from $1$GHz to $10$ GHz or even higher and the strong coupling is about the order of $1$ GHz via mutual inductance or capacitance. The choice of the coupling energy levels are well guaranteed by the rotating wave approximation so long as the large detuning is adjusted. The coupling between the system and a bath can be achieved via resonator and a resistor acts as a bath \cite{ref:coupledqubitsasaswitch,ref:Cottet7561}. In fact, the reservoir could be directly tailored with  the desired bath spectra by reservoir engineering, which was described in detail and applied in many cases \cite{ref:tlmahe,ref:PhysRevE.87.012140,ref:myatt2000decoherence,ref:groblacher2015observation}. In addition, one can note that autonomous quantum refrigerator in a circuit QED architecture based on a Josephson junction and a quantum heat switch based on coupled superconducting qubits have been proposed in Ref. \cite{ref:PhysRevB.94.235420,ref:coupledqubitsasaswitch}, and other relevant investigations about heat transport can also be found in their references.

In conclusion, we have presented a thermal device to realize the functions of a thermal transistor by utilizing the strong internal coupling between the qubit and the qutrit which are connected to three baths with different temperatures. We mainly emphasize the functions as the thermal switch, the modulation, the stabilization, and the amplification which are rigorously demonstrated by both the numerical and the approximately analytic procedures. It is shown that the adjustable energy levels in the qutrit system plays the significant role in the design of the thermal transistor. We also present the possible experimental scheme to realize the scheme.

\section*{ACKNOWLEDGEMENTS}

This work was supported by the National Natural Science Foundation of China, under Grant No.11775040 and No. 11375036, the Xinghai Scholar Cultivation
Plan, and the Fundamental Research Fund for the Central Universities under Grants No. DUT18LK45.

\appendix

\section{Eigenoperators of the system \label{appx:sec:eigenoperators}}

In order to derive the master equation,  we would like to emphasize that the Born-Markov approximation and secular approximation will be used following the standard procedure \cite{ref:ttooqs}. We also require a large internal coupling  $g$ to satisfy the secular approximation condition.  Considering the total Hamiltonian including the three baths as 
\begin{align}
H=H_{S} + H_{SB} + \sum_{\mu} H_{\mu},
\end{align}
we can first diagonalize the Hamiltonian  $H_{S}$ and then in the $H_{S}$ representation derive the eigenoperators with their corresponding eigenfrequencies $\omega_{\mu
l} $ as {\allowdisplaybreaks[4] 
\begin{align}
& V_{L1}= \frac{1}{\sqrt{2}}(\left\vert \lambda_3 \right\rangle \left\langle
\lambda_2 \right\vert),\quad \omega_{L1}=E_2 - g , \\
& V_{L2}= \left\vert \lambda_6 \right\rangle \left\langle \lambda_5
\right\vert,\quad \omega_{L2}=E_2 , \\
& V_{L3}= \frac{1}{\sqrt{2}}(\left\vert \lambda_3 \right\rangle \left\langle
\lambda_4 \right\vert),\quad \omega_{L3}=E_2 + g , \\
& V_{M1}= \left\vert \lambda_6 \right\rangle \left\langle \lambda_3
\right\vert,\quad \omega_{M1}=E_1 , \\
& V_{M2}= \frac{1}{\sqrt{2}}(\left\vert \lambda_5 \right\rangle \left\langle
\lambda_2 \right\vert + \left\vert \lambda_4 \right\rangle \left\langle
\lambda_1 \right\vert),\quad \omega_{M2}=E_1-g, \\
& V_{M3}= \frac{1}{\sqrt{2}}(\left\vert \lambda_5 \right\rangle \left\langle
\lambda_4 \right\vert - \left\vert \lambda_2 \right\rangle \left\langle
\lambda_1 \right\vert),\quad \omega_{M3}=E_1+g, \\
& V_{R1}= \frac{1}{\sqrt{2}}(\left\vert \lambda_6 \right\rangle \left\langle
\lambda_2 \right\vert),\quad \omega_{R1}=E_3 - g , \\
& V_{R2}= -\frac{1}{\sqrt{2}}(\left\vert \lambda_6 \right\rangle
\left\langle \lambda_4 \right\vert),\quad \omega_{R2}=E_3 + g , \\
& V_{R3}= \left\vert \lambda_3 \right\rangle \left\langle \lambda_1
\right\vert,\quad \omega_{R3}=E_3 .
\end{align}
}%
The master equation can be directly obtained by substituting these
eigenoperators into the standard Lindbladian master equation 
\begin{align}
\frac{d \rho_{S}}{dt}&= -i [H_{S}, \rho_{S}] \notag \\
&+ \sum^{N^2 -1}_{k=1} \gamma_{k}
(A_{k}\rho_{S}A^{\dagger}_{k}-\frac{1}{2}A^{\dagger}_{k}A_{k}\rho_{S}-\frac{1%
}{2}\rho_{S}A^{\dagger}_{k}A_{k}).
\end{align}
What we should pay attention to is the $\gamma$'s formula. The master
equation for our model can be found in the main text, we will not show it here
again.
\section{The transistor with different modulation terminals \label{appx:sec:terminals}}

\begin{figure}[H]
\centering
\includegraphics[width=0.75\columnwidth]{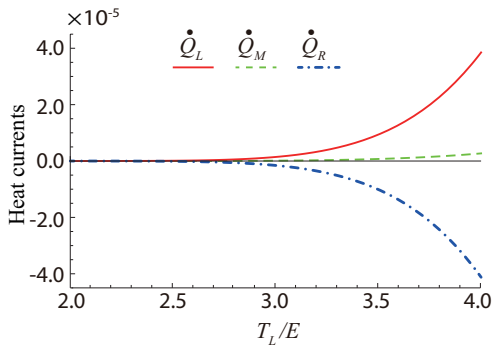}
\caption{(Colour online) Three thermal currents $\dot{Q}_{L}$, $\dot{Q}_{M}$, and $\dot{Q}_{R}$ via numerical method at steady state versus $T_{L}$. Here $T_{M}/E=4$, $T_{R}/E=2$ and the other parameters are same as in Fig.~\ref{fig:populationsTM}. }
\label{fig:transistorcurrentsTL}
\end{figure}

\begin{figure}[H]
\centering
\includegraphics[width=0.75\columnwidth]{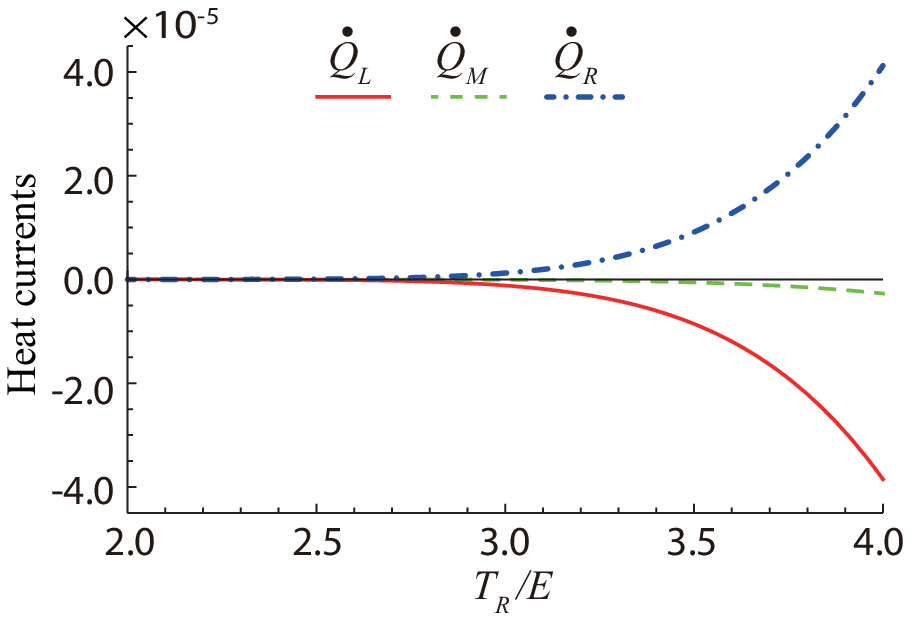}
\caption{(Colour online) Three thermal currents $\dot{Q}_{L}$, $\dot{Q}_{M}$, and $\dot{Q}_{R}$ via numerical method at steady state versus $T_{R}$. Here $T_{M}/E=4$, $T_{L}/E=2$ and the other parameters are same as in Fig.~\ref{fig:populationsTM}. }
\label{fig:transistorcurrentsTR}
\end{figure}

\begin{figure}[H]
\centering
\includegraphics[width=0.75\columnwidth]{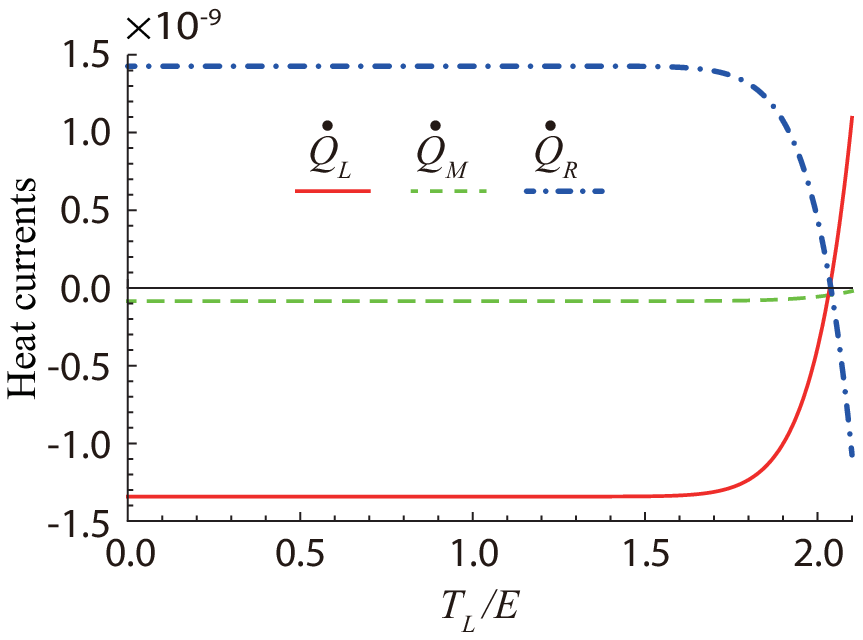}
\caption{(Colour online) Three thermal currents $\dot{Q}_{L}$, $\dot{Q}_{M}$, and $\dot{Q}_{R}$ via numerical method at steady state versus $T_{L}$. Here $T_{M}/E=1.5$, $T_{R}/E=2$ and the other parameters are same as in Fig.~\ref{fig:populationsTM}.}
\label{fig:stablizercurrentsTL}
\end{figure}

In the main text, we consider $\dot{Q}_M$ as the modulation current which can effectively control the heat current between the other two terminals. Here we will show that $\dot{Q}_{L/R}$ can also be used as the modulation current, which is explicitly illustrated in Fig. \ref{fig:transistorcurrentsTL} and Fig. \ref{fig:transistorcurrentsTR}. It is obvious that the functions of interests like the switch, the modulation and the amplification can be realized in the different parameter ranges. In addition, in Fig. \ref{fig:stablizercurrentsTL} we plot the function of the heat current stabilization subject to the temperature fluctuation at the terminal \textit{L}. Although a different lower-temperature terminal is used contrast to the main text, the heat currents are only robust to the temperature fluctuation of the lower-temperature terminal.%

\bibliography{qubit-qutrit-model}

\begin{thebibliography}{78}%
\makeatletter
\providecommand \@ifxundefined [1]{%
 \@ifx{#1\undefined}
}%
\providecommand \@ifnum [1]{%
 \ifnum #1\expandafter \@firstoftwo
 \else \expandafter \@secondoftwo
 \fi
}%
\providecommand \@ifx [1]{%
 \ifx #1\expandafter \@firstoftwo
 \else \expandafter \@secondoftwo
 \fi
}%
\providecommand \natexlab [1]{#1}%
\providecommand \enquote  [1]{``#1''}%
\providecommand \bibnamefont  [1]{#1}%
\providecommand \bibfnamefont [1]{#1}%
\providecommand \citenamefont [1]{#1}%
\providecommand \href@noop [0]{\@secondoftwo}%
\providecommand \href [0]{\begingroup \@sanitize@url \@href}%
\providecommand \@href[1]{\@@startlink{#1}\@@href}%
\providecommand \@@href[1]{\endgroup#1\@@endlink}%
\providecommand \@sanitize@url [0]{\catcode `\\12\catcode `\$12\catcode
  `\&12\catcode `\#12\catcode `\^12\catcode `\_12\catcode `\%12\relax}%
\providecommand \@@startlink[1]{}%
\providecommand \@@endlink[0]{}%
\providecommand \url  [0]{\begingroup\@sanitize@url \@url }%
\providecommand \@url [1]{\endgroup\@href {#1}{\urlprefix }}%
\providecommand \urlprefix  [0]{URL }%
\providecommand \Eprint [0]{\href }%
\providecommand \doibase [0]{http://dx.doi.org/}%
\providecommand \selectlanguage [0]{\@gobble}%
\providecommand \bibinfo  [0]{\@secondoftwo}%
\providecommand \bibfield  [0]{\@secondoftwo}%
\providecommand \translation [1]{[#1]}%
\providecommand \BibitemOpen [0]{}%
\providecommand \bibitemStop [0]{}%
\providecommand \bibitemNoStop [0]{.\EOS\space}%
\providecommand \EOS [0]{\spacefactor3000\relax}%
\providecommand \BibitemShut  [1]{\csname bibitem#1\endcsname}%
\let\auto@bib@innerbib\@empty
%</preamble>
\bibitem [{\citenamefont {Lashkaryov}(1941)}]{ioablbttpm}%
  \BibitemOpen
  \bibfield  {author} {\bibinfo {author} {\bibfnamefont {V.~E.}\ \bibnamefont
  {Lashkaryov}},\ }\bibfield  {title} {\enquote {\bibinfo {title}
  {Investigations of a barrier layer by the thermoprobe method},}\ }\href@noop
  {} {\bibfield  {journal} {\bibinfo  {journal} {Izv. Akad. Nauk SSSR, Ser.
  Fiz.}\ }\textbf {\bibinfo {volume} {5}},\ \bibinfo {pages} {422} (\bibinfo
  {year} {1941})}\BibitemShut {NoStop}%
\bibitem [{\citenamefont {Bardeen}\ and\ \citenamefont
  {Brattain}(1998)}]{658753}%
  \BibitemOpen
  \bibfield  {author} {\bibinfo {author} {\bibfnamefont {J.}~\bibnamefont
  {Bardeen}}\ and\ \bibinfo {author} {\bibfnamefont {W.~H.}\ \bibnamefont
  {Brattain}},\ }\bibfield  {title} {\enquote {\bibinfo {title} {The
  transistor, a semiconductor triode},}\ }\href {\doibase
  10.1109/JPROC.1998.658753} {\bibfield  {journal} {\bibinfo  {journal} {Proc.
  IEEE}\ }\textbf {\bibinfo {volume} {86}},\ \bibinfo {pages} {29} (\bibinfo
  {year} {1998})}\BibitemShut {NoStop}%
\bibitem [{\citenamefont {Chang}\ \emph {et~al.}(2006)\citenamefont {Chang},
  \citenamefont {Okawa}, \citenamefont {Majumdar},\ and\ \citenamefont
  {Zettl}}]{Chang2006Solid}%
  \BibitemOpen
  \bibfield  {author} {\bibinfo {author} {\bibfnamefont {C.~W.}\ \bibnamefont
  {Chang}}, \bibinfo {author} {\bibfnamefont {D.}~\bibnamefont {Okawa}},
  \bibinfo {author} {\bibfnamefont {A.}~\bibnamefont {Majumdar}}, \ and\
  \bibinfo {author} {\bibfnamefont {A.}~\bibnamefont {Zettl}},\ }\bibfield
  {title} {\enquote {\bibinfo {title} {Solid-state thermal rectifier.}}\ }\href
  {http://science.sciencemag.org/content/314/5802/1121} {\bibfield  {journal}
  {\bibinfo  {journal} {Science}\ }\textbf {\bibinfo {volume} {314}},\ \bibinfo
  {pages} {1121} (\bibinfo {year} {2006})}\BibitemShut {NoStop}%
\bibitem [{\citenamefont {Scheibner}\ \emph {et~al.}(2008)\citenamefont
  {Scheibner}, \citenamefont {K\"onig}, \citenamefont {Reuter}, \citenamefont
  {Wieck}, \citenamefont {Gould}, \citenamefont {Buhmann},\ and\ \citenamefont
  {Molenkamp}}]{ref:njp.10.083016}%
  \BibitemOpen
  \bibfield  {author} {\bibinfo {author} {\bibfnamefont {R.}~\bibnamefont
  {Scheibner}}, \bibinfo {author} {\bibfnamefont {M.}~\bibnamefont {K\"onig}},
  \bibinfo {author} {\bibfnamefont {D.}~\bibnamefont {Reuter}}, \bibinfo
  {author} {\bibfnamefont {A.~D.}\ \bibnamefont {Wieck}}, \bibinfo {author}
  {\bibfnamefont {C.}~\bibnamefont {Gould}}, \bibinfo {author} {\bibfnamefont
  {H.}~\bibnamefont {Buhmann}}, \ and\ \bibinfo {author} {\bibfnamefont
  {L.~W.}\ \bibnamefont {Molenkamp}},\ }\bibfield  {title} {\enquote {\bibinfo
  {title} {Quantum dot as thermal rectifier},}\ }\href {\doibase
  10.1088/1367-2630/10/8/083016} {\bibfield  {journal} {\bibinfo  {journal}
  {New J. Phys.}\ }\textbf {\bibinfo {volume} {10}},\ \bibinfo {pages} {083016}
  (\bibinfo {year} {2008})}\BibitemShut {NoStop}%
\bibitem [{\citenamefont {Kobayashi}\ \emph {et~al.}(2009)\citenamefont
  {Kobayashi}, \citenamefont {Teraoka},\ and\ \citenamefont
  {Terasaki}}]{doi:10.1063/1.3253712}%
  \BibitemOpen
  \bibfield  {author} {\bibinfo {author} {\bibfnamefont {W.}~\bibnamefont
  {Kobayashi}}, \bibinfo {author} {\bibfnamefont {Y.}~\bibnamefont {Teraoka}},
  \ and\ \bibinfo {author} {\bibfnamefont {I.}~\bibnamefont {Terasaki}},\
  }\bibfield  {title} {\enquote {\bibinfo {title} {An oxide thermal
  rectifier},}\ }\href {\doibase 10.1063/1.3253712} {\bibfield  {journal}
  {\bibinfo  {journal} {Appl. Phys. Lett.}\ }\textbf {\bibinfo {volume} {95}},\
  \bibinfo {pages} {171905} (\bibinfo {year} {2009})}\BibitemShut {NoStop}%
\bibitem [{\citenamefont {van Zwol}\ \emph
  {et~al.}(2012{\natexlab{a}})\citenamefont {van Zwol}, \citenamefont {Ranno},\
  and\ \citenamefont {Chevrier}}]{van2012Tuning}%
  \BibitemOpen
  \bibfield  {author} {\bibinfo {author} {\bibfnamefont {P.~J.}\ \bibnamefont
  {van Zwol}}, \bibinfo {author} {\bibfnamefont {L.}~\bibnamefont {Ranno}}, \
  and\ \bibinfo {author} {\bibfnamefont {J.}~\bibnamefont {Chevrier}},\
  }\bibfield  {title} {\enquote {\bibinfo {title} {Tuning near field radiative
  heat flux through surface excitations with a metal insulator transition.}}\
  }\href {https://journals.aps.org/prl/abstract/10.1103/PhysRevLett.108.234301}
  {\bibfield  {journal} {\bibinfo  {journal} {Phys. Rev. Lett.}\ }\textbf
  {\bibinfo {volume} {108}},\ \bibinfo {pages} {234301} (\bibinfo {year}
  {2012}{\natexlab{a}})}\BibitemShut {NoStop}%
\bibitem [{\citenamefont {van Zwol}\ \emph
  {et~al.}(2012{\natexlab{b}})\citenamefont {van Zwol}, \citenamefont {Ranno},\
  and\ \citenamefont {Chevrier}}]{doi:10.1063/1.3697673}%
  \BibitemOpen
  \bibfield  {author} {\bibinfo {author} {\bibfnamefont {P.~J.}\ \bibnamefont
  {van Zwol}}, \bibinfo {author} {\bibfnamefont {L.}~\bibnamefont {Ranno}}, \
  and\ \bibinfo {author} {\bibfnamefont {J.}~\bibnamefont {Chevrier}},\
  }\bibfield  {title} {\enquote {\bibinfo {title} {Emissivity measurements with
  an atomic force microscope},}\ }\href {\doibase 10.1063/1.3697673} {\bibfield
   {journal} {\bibinfo  {journal} {J. Appl. Phys.}\ }\textbf {\bibinfo {volume}
  {111}},\ \bibinfo {pages} {063110} (\bibinfo {year}
  {2012}{\natexlab{b}})}\BibitemShut {NoStop}%
\bibitem [{\citenamefont {Ito}\ \emph {et~al.}(2016)\citenamefont {Ito},
  \citenamefont {Nishikawa},\ and\ \citenamefont
  {Iizuka}}]{doi:10.1063/1.4941405}%
  \BibitemOpen
  \bibfield  {author} {\bibinfo {author} {\bibfnamefont {K.}~\bibnamefont
  {Ito}}, \bibinfo {author} {\bibfnamefont {K.}~\bibnamefont {Nishikawa}}, \
  and\ \bibinfo {author} {\bibfnamefont {H.}~\bibnamefont {Iizuka}},\
  }\bibfield  {title} {\enquote {\bibinfo {title} {Multilevel radiative thermal
  memory realized by the hysteretic metal-insulator transition of vanadium
  dioxide},}\ }\href {\doibase 10.1063/1.4941405} {\bibfield  {journal}
  {\bibinfo  {journal} {Appl. Phys. Lett.}\ }\textbf {\bibinfo {volume}
  {108}},\ \bibinfo {pages} {053507} (\bibinfo {year} {2016})}\BibitemShut
  {NoStop}%
\bibitem [{\citenamefont {Ben-Abdallah}\ and\ \citenamefont
  {Biehs}(2013)}]{doi:10.1063/1.4829618}%
  \BibitemOpen
  \bibfield  {author} {\bibinfo {author} {\bibfnamefont {P.}~\bibnamefont
  {Ben-Abdallah}}\ and\ \bibinfo {author} {\bibfnamefont {S.~A.}\ \bibnamefont
  {Biehs}},\ }\bibfield  {title} {\enquote {\bibinfo {title} {Phase-change
  radiative thermal diode},}\ }\href {\doibase 10.1063/1.4829618} {\bibfield
  {journal} {\bibinfo  {journal} {Appl. Phys. Lett.}\ }\textbf {\bibinfo
  {volume} {103}},\ \bibinfo {pages} {191907} (\bibinfo {year}
  {2013})}\BibitemShut {NoStop}%
\bibitem [{\citenamefont {Yang}\ \emph {et~al.}(2013)\citenamefont {Yang},
  \citenamefont {Basu},\ and\ \citenamefont {Wang}}]{doi:10.1063/1.4825168}%
  \BibitemOpen
  \bibfield  {author} {\bibinfo {author} {\bibfnamefont {Y.}~\bibnamefont
  {Yang}}, \bibinfo {author} {\bibfnamefont {S.}~\bibnamefont {Basu}}, \ and\
  \bibinfo {author} {\bibfnamefont {L.~P.}\ \bibnamefont {Wang}},\ }\bibfield
  {title} {\enquote {\bibinfo {title} {Radiation-based near-field thermal
  rectification with phase transition materials},}\ }\href {\doibase
  10.1063/1.4825168} {\bibfield  {journal} {\bibinfo  {journal} {Appl. Phys.
  Lett.}\ }\textbf {\bibinfo {volume} {103}},\ \bibinfo {pages} {163101}
  (\bibinfo {year} {2013})}\BibitemShut {NoStop}%
\bibitem [{\citenamefont {Ito}\ \emph {et~al.}(2014)\citenamefont {Ito},
  \citenamefont {Nishikawa}, \citenamefont {Iizuka},\ and\ \citenamefont
  {Toshiyoshi}}]{doi:10.1063/1.4905132}%
  \BibitemOpen
  \bibfield  {author} {\bibinfo {author} {\bibfnamefont {K.}~\bibnamefont
  {Ito}}, \bibinfo {author} {\bibfnamefont {K.}~\bibnamefont {Nishikawa}},
  \bibinfo {author} {\bibfnamefont {H.}~\bibnamefont {Iizuka}}, \ and\ \bibinfo
  {author} {\bibfnamefont {H.}~\bibnamefont {Toshiyoshi}},\ }\bibfield  {title}
  {\enquote {\bibinfo {title} {Experimental investigation of radiative thermal
  rectifier using vanadium dioxide},}\ }\href {\doibase 10.1063/1.4905132}
  {\bibfield  {journal} {\bibinfo  {journal} {Appl. Phys. Lett.}\ }\textbf
  {\bibinfo {volume} {105}},\ \bibinfo {pages} {253503} (\bibinfo {year}
  {2014})}\BibitemShut {NoStop}%
\bibitem [{\citenamefont {Ben-Abdallah}\ and\ \citenamefont
  {Biehs}(2014)}]{Benabdallah2014Near}%
  \BibitemOpen
  \bibfield  {author} {\bibinfo {author} {\bibfnamefont {P.}~\bibnamefont
  {Ben-Abdallah}}\ and\ \bibinfo {author} {\bibfnamefont {S.~A.}\ \bibnamefont
  {Biehs}},\ }\bibfield  {title} {\enquote {\bibinfo {title} {Near-field
  thermal transistor.}}\ }\href
  {https://doi.org/10.1103/PhysRevLett.112.044301} {\bibfield  {journal}
  {\bibinfo  {journal} {Phys. Rev. Lett.}\ }\textbf {\bibinfo {volume} {112}},\
  \bibinfo {pages} {044301} (\bibinfo {year} {2014})}\BibitemShut {NoStop}%
\bibitem [{\citenamefont {Joulain}\ \emph {et~al.}(2015)\citenamefont
  {Joulain}, \citenamefont {Ezzahri}, \citenamefont {Drevillon},\ and\
  \citenamefont {Ben-Abdallah}}]{doi:10.1063/1.4916730}%
  \BibitemOpen
  \bibfield  {author} {\bibinfo {author} {\bibfnamefont {K.}~\bibnamefont
  {Joulain}}, \bibinfo {author} {\bibfnamefont {Y.}~\bibnamefont {Ezzahri}},
  \bibinfo {author} {\bibfnamefont {J.}~\bibnamefont {Drevillon}}, \ and\
  \bibinfo {author} {\bibfnamefont {P.}~\bibnamefont {Ben-Abdallah}},\
  }\bibfield  {title} {\enquote {\bibinfo {title} {Modulation and amplification
  of radiative far field heat transfer: Towards a simple radiative thermal
  transistor},}\ }\href {\doibase 10.1063/1.4916730} {\bibfield  {journal}
  {\bibinfo  {journal} {Appl. Phys. Lett.}\ }\textbf {\bibinfo {volume}
  {106}},\ \bibinfo {pages} {133505} (\bibinfo {year} {2015})}\BibitemShut
  {NoStop}%
\bibitem [{\citenamefont {Wang}\ \emph {et~al.}(2013)\citenamefont {Wang},
  \citenamefont {Hu},\ and\ \citenamefont {Li}}]{ref:vofliodmclwaii}%
  \BibitemOpen
  \bibfield  {author} {\bibinfo {author} {\bibfnamefont {L.}~\bibnamefont
  {Wang}}, \bibinfo {author} {\bibfnamefont {B.}~\bibnamefont {Hu}}, \ and\
  \bibinfo {author} {\bibfnamefont {B.~W.}\ \bibnamefont {Li}},\ }\bibfield
  {title} {\enquote {\bibinfo {title} {Validity of fourier's law in
  one-dimensional momentum-conserving lattices with asymmetric interparticle
  interactions},}\ }\href {\doibase 10.1103/PhysRevE.88.052112} {\bibfield
  {journal} {\bibinfo  {journal} {Phys. Rev. E}\ }\textbf {\bibinfo {volume}
  {88}},\ \bibinfo {pages} {052112} (\bibinfo {year} {2013})}\BibitemShut
  {NoStop}%
\bibitem [{\citenamefont {Landi}\ and\ \citenamefont
  {de~Oliveira}(2013)}]{ref:flfacocaouecn}%
  \BibitemOpen
  \bibfield  {author} {\bibinfo {author} {\bibfnamefont {G.~T.}\ \bibnamefont
  {Landi}}\ and\ \bibinfo {author} {\bibfnamefont {M.~J.}\ \bibnamefont
  {de~Oliveira}},\ }\bibfield  {title} {\enquote {\bibinfo {title} {Fourier's
  law from a chain of coupled anharmonic oscillators under energy-conserving
  noise},}\ }\href {\doibase 10.1103/PhysRevE.87.052126} {\bibfield  {journal}
  {\bibinfo  {journal} {Phys. Rev. E}\ }\textbf {\bibinfo {volume} {87}},\
  \bibinfo {pages} {052126} (\bibinfo {year} {2013})}\BibitemShut {NoStop}%
\bibitem [{\citenamefont {Chang}\ \emph {et~al.}(2008)\citenamefont {Chang},
  \citenamefont {Okawa}, \citenamefont {Garcia}, \citenamefont {Majumdar},\
  and\ \citenamefont {Zettl}}]{ref:boflintc}%
  \BibitemOpen
  \bibfield  {author} {\bibinfo {author} {\bibfnamefont {C.~W.}\ \bibnamefont
  {Chang}}, \bibinfo {author} {\bibfnamefont {D.}~\bibnamefont {Okawa}},
  \bibinfo {author} {\bibfnamefont {H.}~\bibnamefont {Garcia}}, \bibinfo
  {author} {\bibfnamefont {A.}~\bibnamefont {Majumdar}}, \ and\ \bibinfo
  {author} {\bibfnamefont {A.}~\bibnamefont {Zettl}},\ }\bibfield  {title}
  {\enquote {\bibinfo {title} {Breakdown of fourier's law in nanotube thermal
  conductors},}\ }\href {\doibase 10.1103/PhysRevLett.101.075903} {\bibfield
  {journal} {\bibinfo  {journal} {Phys. Rev. Lett.}\ }\textbf {\bibinfo
  {volume} {101}},\ \bibinfo {pages} {075903} (\bibinfo {year}
  {2008})}\BibitemShut {NoStop}%
\bibitem [{\citenamefont {Manzano}\ \emph {et~al.}(2012)\citenamefont
  {Manzano}, \citenamefont {Tiersch}, \citenamefont {Asadian},\ and\
  \citenamefont {Briegel}}]{ref:qteafl}%
  \BibitemOpen
  \bibfield  {author} {\bibinfo {author} {\bibfnamefont {D.}~\bibnamefont
  {Manzano}}, \bibinfo {author} {\bibfnamefont {M.}~\bibnamefont {Tiersch}},
  \bibinfo {author} {\bibfnamefont {A.}~\bibnamefont {Asadian}}, \ and\
  \bibinfo {author} {\bibfnamefont {H.~J.}\ \bibnamefont {Briegel}},\
  }\bibfield  {title} {\enquote {\bibinfo {title} {Quantum transport efficiency
  and fourier's law},}\ }\href {\doibase 10.1103/PhysRevE.86.061118} {\bibfield
   {journal} {\bibinfo  {journal} {Phys. Rev. E}\ }\textbf {\bibinfo {volume}
  {86}},\ \bibinfo {pages} {061118} (\bibinfo {year} {2012})}\BibitemShut
  {NoStop}%
\bibitem [{\citenamefont {Zhang}\ and\ \citenamefont
  {Zhao}(2002)}]{ref:hciaodas}%
  \BibitemOpen
  \bibfield  {author} {\bibinfo {author} {\bibfnamefont {Y.}~\bibnamefont
  {Zhang}}\ and\ \bibinfo {author} {\bibfnamefont {H.}~\bibnamefont {Zhao}},\
  }\bibfield  {title} {\enquote {\bibinfo {title} {Heat conduction in a
  one-dimensional aperiodic system},}\ }\href {\doibase
  10.1103/PhysRevE.66.026106} {\bibfield  {journal} {\bibinfo  {journal} {Phys.
  Rev. E}\ }\textbf {\bibinfo {volume} {66}},\ \bibinfo {pages} {026106}
  (\bibinfo {year} {2002})}\BibitemShut {NoStop}%
\bibitem [{\citenamefont {Mao}\ \emph {et~al.}(2005)\citenamefont {Mao},
  \citenamefont {Li},\ and\ \citenamefont {Ji}}]{ref:rociodhc}%
  \BibitemOpen
  \bibfield  {author} {\bibinfo {author} {\bibfnamefont {J.~W.}\ \bibnamefont
  {Mao}}, \bibinfo {author} {\bibfnamefont {Y.~Q.}\ \bibnamefont {Li}}, \ and\
  \bibinfo {author} {\bibfnamefont {Y.~Y.}\ \bibnamefont {Ji}},\ }\bibfield
  {title} {\enquote {\bibinfo {title} {Role of chaos in one-dimensional heat
  conductivity},}\ }\href {\doibase 10.1103/PhysRevE.71.061202} {\bibfield
  {journal} {\bibinfo  {journal} {Phys. Rev. E}\ }\textbf {\bibinfo {volume}
  {71}},\ \bibinfo {pages} {061202} (\bibinfo {year} {2005})}\BibitemShut
  {NoStop}%
\bibitem [{\citenamefont {Hu}\ \emph {et~al.}(2006)\citenamefont {Hu},
  \citenamefont {He}, \citenamefont {Yang},\ and\ \citenamefont
  {Zhang}}]{ref:ahctawl}%
  \BibitemOpen
  \bibfield  {author} {\bibinfo {author} {\bibfnamefont {B.}~\bibnamefont
  {Hu}}, \bibinfo {author} {\bibfnamefont {D.}~\bibnamefont {He}}, \bibinfo
  {author} {\bibfnamefont {L.}~\bibnamefont {Yang}}, \ and\ \bibinfo {author}
  {\bibfnamefont {Y.}~\bibnamefont {Zhang}},\ }\bibfield  {title} {\enquote
  {\bibinfo {title} {Asymmetric heat conduction through a weak link},}\ }\href
  {\doibase 10.1103/PhysRevE.74.060101} {\bibfield  {journal} {\bibinfo
  {journal} {Phys. Rev. E}\ }\textbf {\bibinfo {volume} {74}},\ \bibinfo
  {pages} {060101} (\bibinfo {year} {2006})}\BibitemShut {NoStop}%
\bibitem [{\citenamefont {Levy}\ and\ \citenamefont
  {Kosloff}(2014)}]{ref:tlatqtmvtslot}%
  \BibitemOpen
  \bibfield  {author} {\bibinfo {author} {\bibfnamefont {A.}~\bibnamefont
  {Levy}}\ and\ \bibinfo {author} {\bibfnamefont {R.}~\bibnamefont {Kosloff}},\
  }\bibfield  {title} {\enquote {\bibinfo {title} {The local approach to
  quantum transport may violate the second law of thermodynamics},}\ }\href
  {http://stacks.iop.org/0295-5075/107/i=2/a=20004} {\bibfield  {journal}
  {\bibinfo  {journal} {Europhys. Lett.}\ }\textbf {\bibinfo {volume} {107}},\
  \bibinfo {pages} {20004} (\bibinfo {year} {2014})}\BibitemShut {NoStop}%
\bibitem [{\citenamefont {Landsberg}(1956)}]{ref:fot}%
  \BibitemOpen
  \bibfield  {author} {\bibinfo {author} {\bibfnamefont {P.~T.}\ \bibnamefont
  {Landsberg}},\ }\bibfield  {title} {\enquote {\bibinfo {title} {Foundations
  of thermodynamics},}\ }\href {\doibase 10.1103/RevModPhys.28.363} {\bibfield
  {journal} {\bibinfo  {journal} {Rev. Mod. Phys.}\ }\textbf {\bibinfo {volume}
  {28}},\ \bibinfo {pages} {363} (\bibinfo {year} {1956})}\BibitemShut
  {NoStop}%
\bibitem [{\citenamefont {Levy}\ \emph {et~al.}(2012)\citenamefont {Levy},
  \citenamefont {Alicki},\ and\ \citenamefont {Kosloff}}]{ref:qrattlot}%
  \BibitemOpen
  \bibfield  {author} {\bibinfo {author} {\bibfnamefont {A.}~\bibnamefont
  {Levy}}, \bibinfo {author} {\bibfnamefont {R.}~\bibnamefont {Alicki}}, \ and\
  \bibinfo {author} {\bibfnamefont {R.}~\bibnamefont {Kosloff}},\ }\bibfield
  {title} {\enquote {\bibinfo {title} {Quantum refrigerators and the third law
  of thermodynamics},}\ }\href {\doibase 10.1103/PhysRevE.85.061126} {\bibfield
   {journal} {\bibinfo  {journal} {Phys. Rev. E}\ }\textbf {\bibinfo {volume}
  {85}},\ \bibinfo {pages} {061126} (\bibinfo {year} {2012})}\BibitemShut
  {NoStop}%
\bibitem [{\citenamefont {Maruyama}\ \emph {et~al.}(2009)\citenamefont
  {Maruyama}, \citenamefont {Nori},\ and\ \citenamefont
  {Vedral}}]{ref:ctpomdai}%
  \BibitemOpen
  \bibfield  {author} {\bibinfo {author} {\bibfnamefont {K.}~\bibnamefont
  {Maruyama}}, \bibinfo {author} {\bibfnamefont {F.}~\bibnamefont {Nori}}, \
  and\ \bibinfo {author} {\bibfnamefont {V.}~\bibnamefont {Vedral}},\
  }\bibfield  {title} {\enquote {\bibinfo {title} {Colloquium: The physics of
  maxwell’s demon and information},}\ }\href {\doibase
  10.1103/RevModPhys.81.1} {\bibfield  {journal} {\bibinfo  {journal} {Rev.
  Mod. Phys.}\ }\textbf {\bibinfo {volume} {81}},\ \bibinfo {pages} {1}
  (\bibinfo {year} {2009})}\BibitemShut {NoStop}%
\bibitem [{\citenamefont {Feldmann}\ and\ \citenamefont
  {Kosloff}(2000)}]{ref:podheahpift}%
  \BibitemOpen
  \bibfield  {author} {\bibinfo {author} {\bibfnamefont {T.}~\bibnamefont
  {Feldmann}}\ and\ \bibinfo {author} {\bibfnamefont {R.}~\bibnamefont
  {Kosloff}},\ }\bibfield  {title} {\enquote {\bibinfo {title} {Performance of
  discrete heat engines and heat pumps in finite time},}\ }\href {\doibase
  10.1103/PhysRevE.61.4774} {\bibfield  {journal} {\bibinfo  {journal} {Phys.
  Rev. E}\ }\textbf {\bibinfo {volume} {61}},\ \bibinfo {pages} {4774}
  (\bibinfo {year} {2000})}\BibitemShut {NoStop}%
\bibitem [{\citenamefont {Palao}\ \emph {et~al.}(2001)\citenamefont {Palao},
  \citenamefont {Kosloff},\ and\ \citenamefont {Gordon}}]{ref:qtcc}%
  \BibitemOpen
  \bibfield  {author} {\bibinfo {author} {\bibfnamefont {J.~P.}\ \bibnamefont
  {Palao}}, \bibinfo {author} {\bibfnamefont {R.}~\bibnamefont {Kosloff}}, \
  and\ \bibinfo {author} {\bibfnamefont {J.~M.}\ \bibnamefont {Gordon}},\
  }\bibfield  {title} {\enquote {\bibinfo {title} {Quantum thermodynamic
  cooling cycle},}\ }\href {\doibase 10.1103/PhysRevE.64.056130} {\bibfield
  {journal} {\bibinfo  {journal} {Phys. Rev. E}\ }\textbf {\bibinfo {volume}
  {64}},\ \bibinfo {pages} {056130} (\bibinfo {year} {2001})}\BibitemShut
  {NoStop}%
\bibitem [{\citenamefont {Arnaud}\ \emph {et~al.}(2002)\citenamefont {Arnaud},
  \citenamefont {Chusseau},\ and\ \citenamefont {Philippe}}]{ref:ccfao}%
  \BibitemOpen
  \bibfield  {author} {\bibinfo {author} {\bibfnamefont {J.}~\bibnamefont
  {Arnaud}}, \bibinfo {author} {\bibfnamefont {L.}~\bibnamefont {Chusseau}}, \
  and\ \bibinfo {author} {\bibfnamefont {F.}~\bibnamefont {Philippe}},\
  }\bibfield  {title} {\enquote {\bibinfo {title} {Carnot cycle for an
  oscillator},}\ }\href {http://stacks.iop.org/0143-0807/23/i=5/a=306}
  {\bibfield  {journal} {\bibinfo  {journal} {Eur. J. Phys.}\ }\textbf
  {\bibinfo {volume} {23}},\ \bibinfo {pages} {489} (\bibinfo {year}
  {2002})}\BibitemShut {NoStop}%
\bibitem [{\citenamefont {Segal}\ and\ \citenamefont {Nitzan}(2006)}]{ref:mhp}%
  \BibitemOpen
  \bibfield  {author} {\bibinfo {author} {\bibfnamefont {D.}~\bibnamefont
  {Segal}}\ and\ \bibinfo {author} {\bibfnamefont {A.}~\bibnamefont {Nitzan}},\
  }\bibfield  {title} {\enquote {\bibinfo {title} {Molecular heat pump},}\
  }\href {\doibase 10.1103/PhysRevE.73.026109} {\bibfield  {journal} {\bibinfo
  {journal} {Phys. Rev. E}\ }\textbf {\bibinfo {volume} {73}},\ \bibinfo
  {pages} {026109} (\bibinfo {year} {2006})}\BibitemShut {NoStop}%
\bibitem [{\citenamefont {de~Tom\'as}\ \emph {et~al.}(2012)\citenamefont
  {de~Tom\'as}, \citenamefont {Hern\'andez},\ and\ \citenamefont
  {Roco}}]{ref:olsdcear}%
  \BibitemOpen
  \bibfield  {author} {\bibinfo {author} {\bibfnamefont {C.}~\bibnamefont
  {de~Tom\'as}}, \bibinfo {author} {\bibfnamefont {A.~C.}\ \bibnamefont
  {Hern\'andez}}, \ and\ \bibinfo {author} {\bibfnamefont {J.~M.~M.}\
  \bibnamefont {Roco}},\ }\bibfield  {title} {\enquote {\bibinfo {title}
  {Optimal low symmetric dissipation carnot engines and refrigerators},}\
  }\href {\doibase 10.1103/PhysRevE.85.010104} {\bibfield  {journal} {\bibinfo
  {journal} {Phys. Rev. E}\ }\textbf {\bibinfo {volume} {85}},\ \bibinfo
  {pages} {010104} (\bibinfo {year} {2012})}\BibitemShut {NoStop}%
\bibitem [{\citenamefont {Geva}\ and\ \citenamefont
  {Kosloff}(1992)}]{ref:aqmheoift}%
  \BibitemOpen
  \bibfield  {author} {\bibinfo {author} {\bibfnamefont {E.}~\bibnamefont
  {Geva}}\ and\ \bibinfo {author} {\bibfnamefont {R.}~\bibnamefont {Kosloff}},\
  }\bibfield  {title} {\enquote {\bibinfo {title} {A quantum-mechanical heat
  engine operating in finite time. a model consisting of spin-$1/2$ systems as
  the working fluid},}\ }\href {\doibase 10.1063/1.461951} {\bibfield
  {journal} {\bibinfo  {journal} {J. Chem. Phys.}\ }\textbf {\bibinfo {volume}
  {96}},\ \bibinfo {pages} {3054} (\bibinfo {year} {1992})}\BibitemShut
  {NoStop}%
\bibitem [{\citenamefont {Geva}\ and\ \citenamefont
  {Kosloff}(1996)}]{ref:tqheahp}%
  \BibitemOpen
  \bibfield  {author} {\bibinfo {author} {\bibfnamefont {E.}~\bibnamefont
  {Geva}}\ and\ \bibinfo {author} {\bibfnamefont {R.}~\bibnamefont {Kosloff}},\
  }\bibfield  {title} {\enquote {\bibinfo {title} {The quantum heat engine and
  heat pump: An irreversible thermodynamic analysis of the three‐level
  amplifier},}\ }\href {\doibase 10.1063/1.471453} {\bibfield  {journal}
  {\bibinfo  {journal} {J. Chem. Phys.}\ }\textbf {\bibinfo {volume} {104}},\
  \bibinfo {pages} {7681} (\bibinfo {year} {1996})}\BibitemShut {NoStop}%
\bibitem [{\citenamefont {Kosloff}\ and\ \citenamefont
  {Feldmann}(2010)}]{ref:opordqr}%
  \BibitemOpen
  \bibfield  {author} {\bibinfo {author} {\bibfnamefont {R.}~\bibnamefont
  {Kosloff}}\ and\ \bibinfo {author} {\bibfnamefont {T.}~\bibnamefont
  {Feldmann}},\ }\bibfield  {title} {\enquote {\bibinfo {title} {Optimal
  performance of reciprocating demagnetization quantum refrigerators},}\ }\href
  {\doibase 10.1103/PhysRevE.82.011134} {\bibfield  {journal} {\bibinfo
  {journal} {Phys. Rev. E}\ }\textbf {\bibinfo {volume} {82}},\ \bibinfo
  {pages} {011134} (\bibinfo {year} {2010})}\BibitemShut {NoStop}%
\bibitem [{\citenamefont {Thomas}\ and\ \citenamefont
  {Johal}(2011)}]{ref:cqoc}%
  \BibitemOpen
  \bibfield  {author} {\bibinfo {author} {\bibfnamefont {G.}~\bibnamefont
  {Thomas}}\ and\ \bibinfo {author} {\bibfnamefont {R.~S.}\ \bibnamefont
  {Johal}},\ }\bibfield  {title} {\enquote {\bibinfo {title} {Coupled quantum
  otto cycle},}\ }\href {\doibase 10.1103/PhysRevE.83.031135} {\bibfield
  {journal} {\bibinfo  {journal} {Phys. Rev. E}\ }\textbf {\bibinfo {volume}
  {83}},\ \bibinfo {pages} {031135} (\bibinfo {year} {2011})}\BibitemShut
  {NoStop}%
\bibitem [{\citenamefont {Feldmann}\ \emph {et~al.}(1996)\citenamefont
  {Feldmann}, \citenamefont {Geva}, \citenamefont {Kosloff},\ and\
  \citenamefont {Salamon}}]{ref:heiftgbme}%
  \BibitemOpen
  \bibfield  {author} {\bibinfo {author} {\bibfnamefont {T.}~\bibnamefont
  {Feldmann}}, \bibinfo {author} {\bibfnamefont {E.}~\bibnamefont {Geva}},
  \bibinfo {author} {\bibfnamefont {R.}~\bibnamefont {Kosloff}}, \ and\
  \bibinfo {author} {\bibfnamefont {P.}~\bibnamefont {Salamon}},\ }\bibfield
  {title} {\enquote {\bibinfo {title} {Heat engines in finite time governed by
  master equations},}\ }\href {\doibase 10.1119/1.18197} {\bibfield  {journal}
  {\bibinfo  {journal} {Am. J. Phys.}\ }\textbf {\bibinfo {volume} {64}},\
  \bibinfo {pages} {485} (\bibinfo {year} {1996})}\BibitemShut {NoStop}%
\bibitem [{\citenamefont {Feldmann}\ and\ \citenamefont
  {Kosloff}(2003)}]{ref:qfshetoiamwif}%
  \BibitemOpen
  \bibfield  {author} {\bibinfo {author} {\bibfnamefont {T.}~\bibnamefont
  {Feldmann}}\ and\ \bibinfo {author} {\bibfnamefont {R.}~\bibnamefont
  {Kosloff}},\ }\bibfield  {title} {\enquote {\bibinfo {title} {Quantum
  four-stroke heat engine: Thermodynamic observables in a model with intrinsic
  friction},}\ }\href {\doibase 10.1103/PhysRevE.68.016101} {\bibfield
  {journal} {\bibinfo  {journal} {Phys. Rev. E}\ }\textbf {\bibinfo {volume}
  {68}},\ \bibinfo {pages} {016101} (\bibinfo {year} {2003})}\BibitemShut
  {NoStop}%
\bibitem [{\citenamefont {Quan}\ \emph {et~al.}(2007)\citenamefont {Quan},
  \citenamefont {Liu}, \citenamefont {Sun},\ and\ \citenamefont
  {Nori}}]{ref:qtcaqhe}%
  \BibitemOpen
  \bibfield  {author} {\bibinfo {author} {\bibfnamefont {H.~T.}\ \bibnamefont
  {Quan}}, \bibinfo {author} {\bibfnamefont {Y.~X.}\ \bibnamefont {Liu}},
  \bibinfo {author} {\bibfnamefont {C.~P.}\ \bibnamefont {Sun}}, \ and\
  \bibinfo {author} {\bibfnamefont {F.}~\bibnamefont {Nori}},\ }\bibfield
  {title} {\enquote {\bibinfo {title} {Quantum thermodynamic cycles and quantum
  heat engines},}\ }\href {\doibase 10.1103/PhysRevE.76.031105} {\bibfield
  {journal} {\bibinfo  {journal} {Phys. Rev. E}\ }\textbf {\bibinfo {volume}
  {76}},\ \bibinfo {pages} {031105} (\bibinfo {year} {2007})}\BibitemShut
  {NoStop}%
\bibitem [{\citenamefont {Linden}\ \emph {et~al.}(2010)\citenamefont {Linden},
  \citenamefont {Popescu},\ and\ \citenamefont {Skrzypczyk}}]{ref:hsctmbtspr}%
  \BibitemOpen
  \bibfield  {author} {\bibinfo {author} {\bibfnamefont {N.}~\bibnamefont
  {Linden}}, \bibinfo {author} {\bibfnamefont {S.}~\bibnamefont {Popescu}}, \
  and\ \bibinfo {author} {\bibfnamefont {P.}~\bibnamefont {Skrzypczyk}},\
  }\bibfield  {title} {\enquote {\bibinfo {title} {How small can thermal
  machines be? the smallest possible refrigerator},}\ }\href {\doibase
  10.1103/PhysRevLett.105.130401} {\bibfield  {journal} {\bibinfo  {journal}
  {Phys. Rev. Lett.}\ }\textbf {\bibinfo {volume} {105}},\ \bibinfo {pages}
  {130401} (\bibinfo {year} {2010})}\BibitemShut {NoStop}%
\bibitem [{\citenamefont {Yu}\ and\ \citenamefont
  {Zhu}(2014)}]{ref:retscqritscr}%
  \BibitemOpen
  \bibfield  {author} {\bibinfo {author} {\bibfnamefont {C.~S.}\ \bibnamefont
  {Yu}}\ and\ \bibinfo {author} {\bibfnamefont {Q.~Y.}\ \bibnamefont {Zhu}},\
  }\bibfield  {title} {\enquote {\bibinfo {title} {Re-examining the
  self-contained quantum refrigerator in the strong-coupling regime},}\ }\href
  {\doibase 10.1103/PhysRevE.90.052142} {\bibfield  {journal} {\bibinfo
  {journal} {Phys. Rev. E}\ }\textbf {\bibinfo {volume} {90}},\ \bibinfo
  {pages} {052142} (\bibinfo {year} {2014})}\BibitemShut {NoStop}%
\bibitem [{\citenamefont {Man}\ and\ \citenamefont
  {Xia}(2017)}]{ref:PhysRevE.96.012122}%
  \BibitemOpen
  \bibfield  {author} {\bibinfo {author} {\bibfnamefont {Z.~X.}\ \bibnamefont
  {Man}}\ and\ \bibinfo {author} {\bibfnamefont {Y.~J.}\ \bibnamefont {Xia}},\
  }\bibfield  {title} {\enquote {\bibinfo {title} {Smallest quantum thermal
  machine: The effect of strong coupling and distributed thermal tasks},}\
  }\href {\doibase 10.1103/PhysRevE.96.012122} {\bibfield  {journal} {\bibinfo
  {journal} {Phys. Rev. E}\ }\textbf {\bibinfo {volume} {96}},\ \bibinfo
  {pages} {012122} (\bibinfo {year} {2017})}\BibitemShut {NoStop}%
\bibitem [{\citenamefont {Silva}\ \emph {et~al.}(2015)\citenamefont {Silva},
  \citenamefont {Skrzypczyk},\ and\ \citenamefont {Brunner}}]{ref:sqarwrc}%
  \BibitemOpen
  \bibfield  {author} {\bibinfo {author} {\bibfnamefont {R.}~\bibnamefont
  {Silva}}, \bibinfo {author} {\bibfnamefont {P.}~\bibnamefont {Skrzypczyk}}, \
  and\ \bibinfo {author} {\bibfnamefont {N.}~\bibnamefont {Brunner}},\
  }\bibfield  {title} {\enquote {\bibinfo {title} {Small quantum absorption
  refrigerator with reversed couplings},}\ }\href {\doibase
  10.1103/PhysRevE.92.012136} {\bibfield  {journal} {\bibinfo  {journal} {Phys.
  Rev. E}\ }\textbf {\bibinfo {volume} {92}},\ \bibinfo {pages} {012136}
  (\bibinfo {year} {2015})}\BibitemShut {NoStop}%
\bibitem [{\citenamefont {Abah}\ \emph {et~al.}(2012)\citenamefont {Abah},
  \citenamefont {Ro{\ss}nagel}, \citenamefont {Jacob}, \citenamefont {Deffner},
  \citenamefont {Schmidt-Kaler}, \citenamefont {Singer},\ and\ \citenamefont
  {Lutz}}]{ref:siheamp}%
  \BibitemOpen
  \bibfield  {author} {\bibinfo {author} {\bibfnamefont {O.}~\bibnamefont
  {Abah}}, \bibinfo {author} {\bibfnamefont {J.}~\bibnamefont {Ro{\ss}nagel}},
  \bibinfo {author} {\bibfnamefont {G.}~\bibnamefont {Jacob}}, \bibinfo
  {author} {\bibfnamefont {S.}~\bibnamefont {Deffner}}, \bibinfo {author}
  {\bibfnamefont {F.}~\bibnamefont {Schmidt-Kaler}}, \bibinfo {author}
  {\bibfnamefont {K.}~\bibnamefont {Singer}}, \ and\ \bibinfo {author}
  {\bibfnamefont {E.}~\bibnamefont {Lutz}},\ }\bibfield  {title} {\enquote
  {\bibinfo {title} {Single-ion heat engine at maximum power},}\ }\href
  {\doibase 10.1103/PhysRevLett.109.203006} {\bibfield  {journal} {\bibinfo
  {journal} {Phys. Rev. Lett.}\ }\textbf {\bibinfo {volume} {109}},\ \bibinfo
  {pages} {203006} (\bibinfo {year} {2012})}\BibitemShut {NoStop}%
\bibitem [{\citenamefont {Ro{\ss}nagel}\ \emph {et~al.}(2016)\citenamefont
  {Ro{\ss}nagel}, \citenamefont {Dawkins}, \citenamefont {Tolazzi},
  \citenamefont {Abah}, \citenamefont {Lutz}, \citenamefont {Schmidt-Kaler},\
  and\ \citenamefont {Singer}}]{ref:asahe}%
  \BibitemOpen
  \bibfield  {author} {\bibinfo {author} {\bibfnamefont {J.}~\bibnamefont
  {Ro{\ss}nagel}}, \bibinfo {author} {\bibfnamefont {S.~T.}\ \bibnamefont
  {Dawkins}}, \bibinfo {author} {\bibfnamefont {K.~N.}\ \bibnamefont
  {Tolazzi}}, \bibinfo {author} {\bibfnamefont {O.}~\bibnamefont {Abah}},
  \bibinfo {author} {\bibfnamefont {E.}~\bibnamefont {Lutz}}, \bibinfo {author}
  {\bibfnamefont {F.}~\bibnamefont {Schmidt-Kaler}}, \ and\ \bibinfo {author}
  {\bibfnamefont {K.}~\bibnamefont {Singer}},\ }\bibfield  {title} {\enquote
  {\bibinfo {title} {A single-atom heat engine},}\ }\href {\doibase
  10.1126/science.aad6320} {\bibfield  {journal} {\bibinfo  {journal}
  {Science}\ }\textbf {\bibinfo {volume} {352}},\ \bibinfo {pages} {325}
  (\bibinfo {year} {2016})}\BibitemShut {NoStop}%
\bibitem [{\citenamefont {Scovil}\ and\ \citenamefont
  {Schulz-DuBois}(1959)}]{ref:tlmahe}%
  \BibitemOpen
  \bibfield  {author} {\bibinfo {author} {\bibfnamefont {H.~E.~D.}\
  \bibnamefont {Scovil}}\ and\ \bibinfo {author} {\bibfnamefont {E.~O.}\
  \bibnamefont {Schulz-DuBois}},\ }\bibfield  {title} {\enquote {\bibinfo
  {title} {Three-level masers as heat engines},}\ }\href {\doibase
  10.1103/PhysRevLett.2.262} {\bibfield  {journal} {\bibinfo  {journal} {Phys.
  Rev. Lett.}\ }\textbf {\bibinfo {volume} {2}},\ \bibinfo {pages} {262}
  (\bibinfo {year} {1959})}\BibitemShut {NoStop}%
\bibitem [{\citenamefont {Alicki}(1979)}]{ref:tqosaamothe}%
  \BibitemOpen
  \bibfield  {author} {\bibinfo {author} {\bibfnamefont {R.}~\bibnamefont
  {Alicki}},\ }\bibfield  {title} {\enquote {\bibinfo {title} {The quantum open
  system as a model of the heat engine},}\ }\href
  {http://stacks.iop.org/0305-4470/12/i=5/a=007} {\bibfield  {journal}
  {\bibinfo  {journal} {J. Phys. A}\ }\textbf {\bibinfo {volume} {12}},\
  \bibinfo {pages} {L103} (\bibinfo {year} {1979})}\BibitemShut {NoStop}%
\bibitem [{\citenamefont {Skrzypczyk}\ \emph {et~al.}(2011)\citenamefont
  {Skrzypczyk}, \citenamefont {Brunner}, \citenamefont {Linden},\ and\
  \citenamefont {Popescu}}]{ref:tsrcrme}%
  \BibitemOpen
  \bibfield  {author} {\bibinfo {author} {\bibfnamefont {P.}~\bibnamefont
  {Skrzypczyk}}, \bibinfo {author} {\bibfnamefont {N.}~\bibnamefont {Brunner}},
  \bibinfo {author} {\bibfnamefont {N.}~\bibnamefont {Linden}}, \ and\ \bibinfo
  {author} {\bibfnamefont {S.}~\bibnamefont {Popescu}},\ }\bibfield  {title}
  {\enquote {\bibinfo {title} {The smallest refrigerators can reach maximal
  efficiency},}\ }\href {http://stacks.iop.org/1751-8121/44/i=49/a=492002}
  {\bibfield  {journal} {\bibinfo  {journal} {J. Phys. A}\ }\textbf {\bibinfo
  {volume} {44}},\ \bibinfo {pages} {492002} (\bibinfo {year}
  {2011})}\BibitemShut {NoStop}%
\bibitem [{\citenamefont {Wang}\ and\ \citenamefont {Li}(2007)}]{ref:tlgcwp}%
  \BibitemOpen
  \bibfield  {author} {\bibinfo {author} {\bibfnamefont {L.}~\bibnamefont
  {Wang}}\ and\ \bibinfo {author} {\bibfnamefont {B.~W.}\ \bibnamefont {Li}},\
  }\bibfield  {title} {\enquote {\bibinfo {title} {Thermal logic gates:
  Computation with phonons},}\ }\href {\doibase 10.1103/PhysRevLett.99.177208}
  {\bibfield  {journal} {\bibinfo  {journal} {Phys. Rev. Lett.}\ }\textbf
  {\bibinfo {volume} {99}},\ \bibinfo {pages} {177208} (\bibinfo {year}
  {2007})}\BibitemShut {NoStop}%
\bibitem [{\citenamefont {Wang}\ and\ \citenamefont {Li}(2008)}]{ref:tmasopi}%
  \BibitemOpen
  \bibfield  {author} {\bibinfo {author} {\bibfnamefont {L.}~\bibnamefont
  {Wang}}\ and\ \bibinfo {author} {\bibfnamefont {B.~W.}\ \bibnamefont {Li}},\
  }\bibfield  {title} {\enquote {\bibinfo {title} {Thermal memory: A storage of
  phononic information},}\ }\href {\doibase 10.1103/PhysRevLett.101.267203}
  {\bibfield  {journal} {\bibinfo  {journal} {Phys. Rev. Lett.}\ }\textbf
  {\bibinfo {volume} {101}},\ \bibinfo {pages} {267203} (\bibinfo {year}
  {2008})}\BibitemShut {NoStop}%
\bibitem [{\citenamefont {Faucheux}\ \emph {et~al.}(1995)\citenamefont
  {Faucheux}, \citenamefont {Bourdieu}, \citenamefont {Kaplan},\ and\
  \citenamefont {Libchaber}}]{ref:otr}%
  \BibitemOpen
  \bibfield  {author} {\bibinfo {author} {\bibfnamefont {L.~P.}\ \bibnamefont
  {Faucheux}}, \bibinfo {author} {\bibfnamefont {L.~S.}\ \bibnamefont
  {Bourdieu}}, \bibinfo {author} {\bibfnamefont {P.~D.}\ \bibnamefont
  {Kaplan}}, \ and\ \bibinfo {author} {\bibfnamefont {A.~J.}\ \bibnamefont
  {Libchaber}},\ }\bibfield  {title} {\enquote {\bibinfo {title} {Optical
  thermal ratchet},}\ }\href {\doibase 10.1103/PhysRevLett.74.1504} {\bibfield
  {journal} {\bibinfo  {journal} {Phys. Rev. Lett.}\ }\textbf {\bibinfo
  {volume} {74}},\ \bibinfo {pages} {1504} (\bibinfo {year}
  {1995})}\BibitemShut {NoStop}%
\bibitem [{\citenamefont {Zhan}\ \emph {et~al.}(2009)\citenamefont {Zhan},
  \citenamefont {Li}, \citenamefont {Kohler},\ and\ \citenamefont
  {H\"anggi}}]{ref:mwaaqhr}%
  \BibitemOpen
  \bibfield  {author} {\bibinfo {author} {\bibfnamefont {F.}~\bibnamefont
  {Zhan}}, \bibinfo {author} {\bibfnamefont {N.~B.}\ \bibnamefont {Li}},
  \bibinfo {author} {\bibfnamefont {S.}~\bibnamefont {Kohler}}, \ and\ \bibinfo
  {author} {\bibfnamefont {P.}~\bibnamefont {H\"anggi}},\ }\bibfield  {title}
  {\enquote {\bibinfo {title} {Molecular wires acting as quantum heat
  ratchets},}\ }\href {\doibase 10.1103/PhysRevE.80.061115} {\bibfield
  {journal} {\bibinfo  {journal} {Phys. Rev. E}\ }\textbf {\bibinfo {volume}
  {80}},\ \bibinfo {pages} {061115} (\bibinfo {year} {2009})}\BibitemShut
  {NoStop}%
\bibitem [{\citenamefont {Hofer}\ \emph {et~al.}(2017)\citenamefont {Hofer},
  \citenamefont {Brask}, \citenamefont {Perarnau-Llobet},\ and\ \citenamefont
  {Brunner}}]{ref:PhysRevLett.119.090603}%
  \BibitemOpen
  \bibfield  {author} {\bibinfo {author} {\bibfnamefont {P.~P.}\ \bibnamefont
  {Hofer}}, \bibinfo {author} {\bibfnamefont {J.~B.}\ \bibnamefont {Brask}},
  \bibinfo {author} {\bibfnamefont {M.}~\bibnamefont {Perarnau-Llobet}}, \ and\
  \bibinfo {author} {\bibfnamefont {N.}~\bibnamefont {Brunner}},\ }\bibfield
  {title} {\enquote {\bibinfo {title} {Quantum thermal machine as a
  thermometer},}\ }\href {\doibase 10.1103/PhysRevLett.119.090603} {\bibfield
  {journal} {\bibinfo  {journal} {Phys. Rev. Lett.}\ }\textbf {\bibinfo
  {volume} {119}},\ \bibinfo {pages} {090603} (\bibinfo {year}
  {2017})}\BibitemShut {NoStop}%
\bibitem [{\citenamefont {Werlang}\ \emph {et~al.}(2014)\citenamefont
  {Werlang}, \citenamefont {Marchiori}, \citenamefont {Cornelio},\ and\
  \citenamefont {Valente}}]{ref:oritucr}%
  \BibitemOpen
  \bibfield  {author} {\bibinfo {author} {\bibfnamefont {T.}~\bibnamefont
  {Werlang}}, \bibinfo {author} {\bibfnamefont {M.~A.}\ \bibnamefont
  {Marchiori}}, \bibinfo {author} {\bibfnamefont {M.~F.}\ \bibnamefont
  {Cornelio}}, \ and\ \bibinfo {author} {\bibfnamefont {D.}~\bibnamefont
  {Valente}},\ }\bibfield  {title} {\enquote {\bibinfo {title} {Optimal
  rectification in the ultrastrong coupling regime},}\ }\href {\doibase
  10.1103/PhysRevE.89.062109} {\bibfield  {journal} {\bibinfo  {journal} {Phys.
  Rev. E}\ }\textbf {\bibinfo {volume} {89}},\ \bibinfo {pages} {062109}
  (\bibinfo {year} {2014})}\BibitemShut {NoStop}%
\bibitem [{\citenamefont {Chen}\ and\ \citenamefont
  {Wang}(2015)}]{ref:tritnqds}%
  \BibitemOpen
  \bibfield  {author} {\bibinfo {author} {\bibfnamefont {T.}~\bibnamefont
  {Chen}}\ and\ \bibinfo {author} {\bibfnamefont {X.~B.}\ \bibnamefont
  {Wang}},\ }\bibfield  {title} {\enquote {\bibinfo {title} {Thermal
  rectification in the nonequilibrium quantum-dots-system},}\ }\href {\doibase
  http://dx.doi.org/10.1016/j.physe.2015.04.021} {\bibfield  {journal}
  {\bibinfo  {journal} {Physica E}\ }\textbf {\bibinfo {volume} {72}},\
  \bibinfo {pages} {58} (\bibinfo {year} {2015})}\BibitemShut {NoStop}%
\bibitem [{\citenamefont {Li}\ \emph {et~al.}(2004)\citenamefont {Li},
  \citenamefont {Wang},\ and\ \citenamefont {Casati}}]{ref:tdrohf}%
  \BibitemOpen
  \bibfield  {author} {\bibinfo {author} {\bibfnamefont {B.~W.}\ \bibnamefont
  {Li}}, \bibinfo {author} {\bibfnamefont {L.}~\bibnamefont {Wang}}, \ and\
  \bibinfo {author} {\bibfnamefont {G.}~\bibnamefont {Casati}},\ }\bibfield
  {title} {\enquote {\bibinfo {title} {Thermal diode: Rectification of heat
  flux},}\ }\href {\doibase 10.1103/PhysRevLett.93.184301} {\bibfield
  {journal} {\bibinfo  {journal} {Phys. Rev. Lett.}\ }\textbf {\bibinfo
  {volume} {93}},\ \bibinfo {pages} {184301} (\bibinfo {year}
  {2004})}\BibitemShut {NoStop}%
\bibitem [{\citenamefont {Pereira}(2011)}]{ref:scftriggm}%
  \BibitemOpen
  \bibfield  {author} {\bibinfo {author} {\bibfnamefont {E.}~\bibnamefont
  {Pereira}},\ }\bibfield  {title} {\enquote {\bibinfo {title} {Sufficient
  conditions for thermal rectification in general graded materials},}\ }\href
  {\doibase 10.1103/PhysRevE.83.031106} {\bibfield  {journal} {\bibinfo
  {journal} {Phys. Rev. E}\ }\textbf {\bibinfo {volume} {83}},\ \bibinfo
  {pages} {031106} (\bibinfo {year} {2011})}\BibitemShut {NoStop}%
\bibitem [{\citenamefont {Wang}\ \emph {et~al.}(2012)\citenamefont {Wang},
  \citenamefont {Pereira},\ and\ \citenamefont {Casati}}]{ref:trigm}%
  \BibitemOpen
  \bibfield  {author} {\bibinfo {author} {\bibfnamefont {J.}~\bibnamefont
  {Wang}}, \bibinfo {author} {\bibfnamefont {E.}~\bibnamefont {Pereira}}, \
  and\ \bibinfo {author} {\bibfnamefont {G.}~\bibnamefont {Casati}},\
  }\bibfield  {title} {\enquote {\bibinfo {title} {Thermal rectification in
  graded materials},}\ }\href {\doibase 10.1103/PhysRevE.86.010101} {\bibfield
  {journal} {\bibinfo  {journal} {Phys. Rev. E}\ }\textbf {\bibinfo {volume}
  {86}},\ \bibinfo {pages} {010101} (\bibinfo {year} {2012})}\BibitemShut
  {NoStop}%
\bibitem [{\citenamefont {Fratini}\ and\ \citenamefont
  {Ghobadi}(2016)}]{ref:fqtoald}%
  \BibitemOpen
  \bibfield  {author} {\bibinfo {author} {\bibfnamefont {F.}~\bibnamefont
  {Fratini}}\ and\ \bibinfo {author} {\bibfnamefont {R.}~\bibnamefont
  {Ghobadi}},\ }\bibfield  {title} {\enquote {\bibinfo {title} {Full quantum
  treatment of a light diode},}\ }\href {\doibase 10.1103/PhysRevA.93.023818}
  {\bibfield  {journal} {\bibinfo  {journal} {Phys. Rev. A}\ }\textbf {\bibinfo
  {volume} {93}},\ \bibinfo {pages} {023818} (\bibinfo {year}
  {2016})}\BibitemShut {NoStop}%
\bibitem [{\citenamefont {Landi}\ \emph {et~al.}(2014)\citenamefont {Landi},
  \citenamefont {Novais}, \citenamefont {de~Oliveira},\ and\ \citenamefont
  {Karevski}}]{ref:fritqxxzc}%
  \BibitemOpen
  \bibfield  {author} {\bibinfo {author} {\bibfnamefont {G.~T.}\ \bibnamefont
  {Landi}}, \bibinfo {author} {\bibfnamefont {E.}~\bibnamefont {Novais}},
  \bibinfo {author} {\bibfnamefont {M.~J.}\ \bibnamefont {de~Oliveira}}, \ and\
  \bibinfo {author} {\bibfnamefont {D.}~\bibnamefont {Karevski}},\ }\bibfield
  {title} {\enquote {\bibinfo {title} {Flux rectification in the quantum $xxz$
  chain},}\ }\href {\doibase 10.1103/PhysRevE.90.042142} {\bibfield  {journal}
  {\bibinfo  {journal} {Phys. Rev. E}\ }\textbf {\bibinfo {volume} {90}},\
  \bibinfo {pages} {042142} (\bibinfo {year} {2014})}\BibitemShut {NoStop}%
\bibitem [{\citenamefont {Man}\ \emph {et~al.}(2016)\citenamefont {Man},
  \citenamefont {An},\ and\ \citenamefont {Xia}}]{ref:chfatracttls}%
  \BibitemOpen
  \bibfield  {author} {\bibinfo {author} {\bibfnamefont {Z.~X.}\ \bibnamefont
  {Man}}, \bibinfo {author} {\bibfnamefont {N.~B.}\ \bibnamefont {An}}, \ and\
  \bibinfo {author} {\bibfnamefont {Y.~J.}\ \bibnamefont {Xia}},\ }\bibfield
  {title} {\enquote {\bibinfo {title} {Controlling heat flows among three
  reservoirs asymmetrically coupled to two two-level systems},}\ }\href
  {\doibase 10.1103/PhysRevE.94.042135} {\bibfield  {journal} {\bibinfo
  {journal} {Phys. Rev. E}\ }\textbf {\bibinfo {volume} {94}},\ \bibinfo
  {pages} {042135} (\bibinfo {year} {2016})}\BibitemShut {NoStop}%
\bibitem [{\citenamefont {Jiang}\ \emph {et~al.}(2015)\citenamefont {Jiang},
  \citenamefont {Kulkarni}, \citenamefont {Segal},\ and\ \citenamefont
  {Imry}}]{ref:pttar}%
  \BibitemOpen
  \bibfield  {author} {\bibinfo {author} {\bibfnamefont {J.~H.}\ \bibnamefont
  {Jiang}}, \bibinfo {author} {\bibfnamefont {M.}~\bibnamefont {Kulkarni}},
  \bibinfo {author} {\bibfnamefont {D.}~\bibnamefont {Segal}}, \ and\ \bibinfo
  {author} {\bibfnamefont {Y.}~\bibnamefont {Imry}},\ }\bibfield  {title}
  {\enquote {\bibinfo {title} {Phonon thermoelectric transistors and
  rectifiers},}\ }\href {\doibase 10.1103/PhysRevB.92.045309} {\bibfield
  {journal} {\bibinfo  {journal} {Phys. Rev. B}\ }\textbf {\bibinfo {volume}
  {92}},\ \bibinfo {pages} {045309} (\bibinfo {year} {2015})}\BibitemShut
  {NoStop}%
\bibitem [{\citenamefont {Joulain}\ \emph {et~al.}(2016)\citenamefont
  {Joulain}, \citenamefont {Drevillon}, \citenamefont {Ezzahri},\ and\
  \citenamefont {Ordonez-Miranda}}]{ref:qtt}%
  \BibitemOpen
  \bibfield  {author} {\bibinfo {author} {\bibfnamefont {K.}~\bibnamefont
  {Joulain}}, \bibinfo {author} {\bibfnamefont {J.}~\bibnamefont {Drevillon}},
  \bibinfo {author} {\bibfnamefont {Y.}~\bibnamefont {Ezzahri}}, \ and\
  \bibinfo {author} {\bibfnamefont {J.}~\bibnamefont {Ordonez-Miranda}},\
  }\bibfield  {title} {\enquote {\bibinfo {title} {Quantum thermal
  transistor},}\ }\href {\doibase 10.1103/PhysRevLett.116.200601} {\bibfield
  {journal} {\bibinfo  {journal} {Phys. Rev. Lett.}\ }\textbf {\bibinfo
  {volume} {116}},\ \bibinfo {pages} {200601} (\bibinfo {year}
  {2016})}\BibitemShut {NoStop}%
\bibitem [{\citenamefont {Lo}\ \emph {et~al.}(2008)\citenamefont {Lo},
  \citenamefont {Wang},\ and\ \citenamefont {Li}}]{ref:tthfsam}%
  \BibitemOpen
  \bibfield  {author} {\bibinfo {author} {\bibfnamefont {W.~C.}\ \bibnamefont
  {Lo}}, \bibinfo {author} {\bibfnamefont {L.}~\bibnamefont {Wang}}, \ and\
  \bibinfo {author} {\bibfnamefont {B.~W.}\ \bibnamefont {Li}},\ }\bibfield
  {title} {\enquote {\bibinfo {title} {Thermal transistor: Heat flux switching
  and modulating},}\ }\href {\doibase 10.1143/JPSJ.77.054402} {\bibfield
  {journal} {\bibinfo  {journal} {J. Phys. Soc. Jpn.}\ }\textbf {\bibinfo
  {volume} {77}},\ \bibinfo {pages} {054402} (\bibinfo {year}
  {2008})}\BibitemShut {NoStop}%
\bibitem [{\citenamefont {Li}\ \emph {et~al.}(2006)\citenamefont {Li},
  \citenamefont {Wang},\ and\ \citenamefont {Casati}}]{Li2006Negative}%
  \BibitemOpen
  \bibfield  {author} {\bibinfo {author} {\bibfnamefont {B.~W.}\ \bibnamefont
  {Li}}, \bibinfo {author} {\bibfnamefont {L.}~\bibnamefont {Wang}}, \ and\
  \bibinfo {author} {\bibfnamefont {G.}~\bibnamefont {Casati}},\ }\bibfield
  {title} {\enquote {\bibinfo {title} {Negative differential thermal resistance
  and thermal transistor},}\ }\href {\doibase 10.1063/1.2191730} {\bibfield
  {journal} {\bibinfo  {journal} {Appl. Phys. Lett.}\ }\textbf {\bibinfo
  {volume} {88}},\ \bibinfo {pages} {143501} (\bibinfo {year}
  {2006})}\BibitemShut {NoStop}%
\bibitem [{\citenamefont {Komatsu}\ and\ \citenamefont
  {Ito}(2011)}]{Komatsu2011Thermal}%
  \BibitemOpen
  \bibfield  {author} {\bibinfo {author} {\bibfnamefont {T.~S.}\ \bibnamefont
  {Komatsu}}\ and\ \bibinfo {author} {\bibfnamefont {N.}~\bibnamefont {Ito}},\
  }\bibfield  {title} {\enquote {\bibinfo {title} {Thermal transistor utilizing
  gas-liquid transition},}\ }\href {\doibase 10.1103/PhysRevE.83.012104}
  {\bibfield  {journal} {\bibinfo  {journal} {Phys. Rev. E}\ }\textbf {\bibinfo
  {volume} {83}},\ \bibinfo {pages} {012104} (\bibinfo {year}
  {2011})}\BibitemShut {NoStop}%
\bibitem [{\citenamefont {Joulain}\ \emph {et~al.}(2017)\citenamefont
  {Joulain}, \citenamefont {Ezzahri},\ and\ \citenamefont
  {Ordonez-Miranda}}]{ref:zna.72.163}%
  \BibitemOpen
  \bibfield  {author} {\bibinfo {author} {\bibfnamefont {K.}~\bibnamefont
  {Joulain}}, \bibinfo {author} {\bibfnamefont {Y.}~\bibnamefont {Ezzahri}}, \
  and\ \bibinfo {author} {\bibfnamefont {J.}~\bibnamefont {Ordonez-Miranda}},\
  }\bibfield  {title} {\enquote {\bibinfo {title} {Quantum thermal
  rectification to design thermal diodes and transistors},}\ }\href {\doibase
  10.1515/zna-2016-0350} {\bibfield  {journal} {\bibinfo  {journal} {Z.
  Naturforsch. A}\ }\textbf {\bibinfo {volume} {72}},\ \bibinfo {pages} {163}
  (\bibinfo {year} {2017})}\BibitemShut {NoStop}%
\bibitem [{\citenamefont {You}\ and\ \citenamefont
  {Nori}(2011)}]{ref:you2011atomic}%
  \BibitemOpen
  \bibfield  {author} {\bibinfo {author} {\bibfnamefont {JQ}~\bibnamefont
  {You}}\ and\ \bibinfo {author} {\bibfnamefont {F.}~\bibnamefont {Nori}},\
  }\bibfield  {title} {\enquote {\bibinfo {title} {Atomic physics and quantum
  optics using superconducting circuits},}\ }\href {\doibase
  10.1038/nature10122} {\bibfield  {journal} {\bibinfo  {journal} {Nature}\
  }\textbf {\bibinfo {volume} {474}},\ \bibinfo {pages} {589} (\bibinfo {year}
  {2011})}\BibitemShut {NoStop}%
\bibitem [{\citenamefont {Buluta}\ \emph {et~al.}(2011)\citenamefont {Buluta},
  \citenamefont {Ashhab},\ and\ \citenamefont
  {Nori}}]{ref:naturalandartificialatoms}%
  \BibitemOpen
  \bibfield  {author} {\bibinfo {author} {\bibfnamefont {I.}~\bibnamefont
  {Buluta}}, \bibinfo {author} {\bibfnamefont {S.}~\bibnamefont {Ashhab}}, \
  and\ \bibinfo {author} {\bibfnamefont {F.}~\bibnamefont {Nori}},\ }\bibfield
  {title} {\enquote {\bibinfo {title} {Natural and artificial atoms for quantum
  computation},}\ }\href {\doibase 10.1088/0034-4885/74/10/104401} {\bibfield
  {journal} {\bibinfo  {journal} {Reports on Progress in Physics}\ }\textbf
  {\bibinfo {volume} {74}},\ \bibinfo {pages} {104401} (\bibinfo {year}
  {2011})}\BibitemShut {NoStop}%
\bibitem [{\citenamefont {Hofer}\ \emph {et~al.}(2016)\citenamefont {Hofer},
  \citenamefont {Perarnau-Llobet}, \citenamefont {Brask}, \citenamefont
  {Silva}, \citenamefont {Huber},\ and\ \citenamefont
  {Brunner}}]{ref:PhysRevB.94.235420}%
  \BibitemOpen
  \bibfield  {author} {\bibinfo {author} {\bibfnamefont {P.~P.}\ \bibnamefont
  {Hofer}}, \bibinfo {author} {\bibfnamefont {M.}~\bibnamefont
  {Perarnau-Llobet}}, \bibinfo {author} {\bibfnamefont {J.~B.}\ \bibnamefont
  {Brask}}, \bibinfo {author} {\bibfnamefont {R.}~\bibnamefont {Silva}},
  \bibinfo {author} {\bibfnamefont {M.}~\bibnamefont {Huber}}, \ and\ \bibinfo
  {author} {\bibfnamefont {N.}~\bibnamefont {Brunner}},\ }\bibfield  {title}
  {\enquote {\bibinfo {title} {Autonomous quantum refrigerator in a circuit qed
  architecture based on a josephson junction},}\ }\href {\doibase
  10.1103/PhysRevB.94.235420} {\bibfield  {journal} {\bibinfo  {journal} {Phys.
  Rev. B}\ }\textbf {\bibinfo {volume} {94}},\ \bibinfo {pages} {235420}
  (\bibinfo {year} {2016})}\BibitemShut {NoStop}%
\bibitem [{\citenamefont {Karimi}\ \emph {et~al.}(2017)\citenamefont {Karimi},
  \citenamefont {Pekola}, \citenamefont {Campisi},\ and\ \citenamefont
  {Fazio}}]{ref:coupledqubitsasaswitch}%
  \BibitemOpen
  \bibfield  {author} {\bibinfo {author} {\bibfnamefont {B.}~\bibnamefont
  {Karimi}}, \bibinfo {author} {\bibfnamefont {J.~P.}\ \bibnamefont {Pekola}},
  \bibinfo {author} {\bibfnamefont {M.}~\bibnamefont {Campisi}}, \ and\
  \bibinfo {author} {\bibfnamefont {R.}~\bibnamefont {Fazio}},\ }\bibfield
  {title} {\enquote {\bibinfo {title} {Coupled qubits as a quantum heat
  switch},}\ }\href {\doibase 10.1088/2058-9565/aa8330} {\bibfield  {journal}
  {\bibinfo  {journal} {Quantum Science and Technology}\ }\textbf {\bibinfo
  {volume} {2}},\ \bibinfo {pages} {044007} (\bibinfo {year}
  {2017})}\BibitemShut {NoStop}%
\bibitem [{\citenamefont {Breuer}\ and\ \citenamefont
  {Petruccione}(2002)}]{ref:ttooqs}%
  \BibitemOpen
  \bibfield  {author} {\bibinfo {author} {\bibfnamefont {H.~P.}\ \bibnamefont
  {Breuer}}\ and\ \bibinfo {author} {\bibfnamefont {F.}~\bibnamefont
  {Petruccione}},\ }\href@noop {} {\emph {\bibinfo {title} {The Theory of Open
  Quantum Systems}}}\ (\bibinfo  {publisher} {Oxford University Press},\
  \bibinfo {address} {Oxford, UK},\ \bibinfo {year} {2002})\BibitemShut
  {NoStop}%
\bibitem [{\citenamefont {Szczygielski}\ \emph {et~al.}(2013)\citenamefont
  {Szczygielski}, \citenamefont {Gelbwaser-Klimovsky},\ and\ \citenamefont
  {Alicki}}]{ref:PhysRevE.87.012120}%
  \BibitemOpen
  \bibfield  {author} {\bibinfo {author} {\bibfnamefont {K.}~\bibnamefont
  {Szczygielski}}, \bibinfo {author} {\bibfnamefont {D.}~\bibnamefont
  {Gelbwaser-Klimovsky}}, \ and\ \bibinfo {author} {\bibfnamefont
  {R.}~\bibnamefont {Alicki}},\ }\bibfield  {title} {\enquote {\bibinfo {title}
  {Markovian master equation and thermodynamics of a two-level system in a
  strong laser field},}\ }\href {\doibase 10.1103/PhysRevE.87.012120}
  {\bibfield  {journal} {\bibinfo  {journal} {Phys. Rev. E}\ }\textbf {\bibinfo
  {volume} {87}},\ \bibinfo {pages} {012120} (\bibinfo {year}
  {2013})}\BibitemShut {NoStop}%
\bibitem [{\citenamefont {Alicki}\ \emph {et~al.}(2006)\citenamefont {Alicki},
  \citenamefont {Lidar},\ and\ \citenamefont
  {Zanardi}}]{ref:PhysRevA.73.052311}%
  \BibitemOpen
  \bibfield  {author} {\bibinfo {author} {\bibfnamefont {R.}~\bibnamefont
  {Alicki}}, \bibinfo {author} {\bibfnamefont {D.~A.}\ \bibnamefont {Lidar}}, \
  and\ \bibinfo {author} {\bibfnamefont {P.}~\bibnamefont {Zanardi}},\
  }\bibfield  {title} {\enquote {\bibinfo {title} {Internal consistency of
  fault-tolerant quantum error correction in light of rigorous derivations of
  the quantum markovian limit},}\ }\href {\doibase 10.1103/PhysRevA.73.052311}
  {\bibfield  {journal} {\bibinfo  {journal} {Phys. Rev. A}\ }\textbf {\bibinfo
  {volume} {73}},\ \bibinfo {pages} {052311} (\bibinfo {year}
  {2006})}\BibitemShut {NoStop}%
\bibitem [{\citenamefont {Kol\'a\ifmmode~\check{r}\else \v{r}\fi{}}\ \emph
  {et~al.}(2012)\citenamefont {Kol\'a\ifmmode~\check{r}\else \v{r}\fi{}},
  \citenamefont {Gelbwaser-Klimovsky}, \citenamefont {Alicki},\ and\
  \citenamefont {Kurizki}}]{ref:PhysRevLett.109.090601}%
  \BibitemOpen
  \bibfield  {author} {\bibinfo {author} {\bibfnamefont {M.}~\bibnamefont
  {Kol\'a\ifmmode~\check{r}\else \v{r}\fi{}}}, \bibinfo {author} {\bibfnamefont
  {D.}~\bibnamefont {Gelbwaser-Klimovsky}}, \bibinfo {author} {\bibfnamefont
  {R.}~\bibnamefont {Alicki}}, \ and\ \bibinfo {author} {\bibfnamefont
  {G.}~\bibnamefont {Kurizki}},\ }\bibfield  {title} {\enquote {\bibinfo
  {title} {Quantum bath refrigeration towards absolute zero: Challenging the
  unattainability principle},}\ }\href {\doibase
  10.1103/PhysRevLett.109.090601} {\bibfield  {journal} {\bibinfo  {journal}
  {Phys. Rev. Lett.}\ }\textbf {\bibinfo {volume} {109}},\ \bibinfo {pages}
  {090601} (\bibinfo {year} {2012})}\BibitemShut {NoStop}%
\bibitem [{\citenamefont {Niskanen}\ \emph {et~al.}(2007)\citenamefont
  {Niskanen}, \citenamefont {Harrabi}, \citenamefont {Yoshihara}, \citenamefont
  {Nakamura}, \citenamefont {Lloyd},\ and\ \citenamefont
  {Tsai}}]{ref:niskanen2007quantum}%
  \BibitemOpen
  \bibfield  {author} {\bibinfo {author} {\bibfnamefont {A.~O.}\ \bibnamefont
  {Niskanen}}, \bibinfo {author} {\bibfnamefont {K.}~\bibnamefont {Harrabi}},
  \bibinfo {author} {\bibfnamefont {F.}~\bibnamefont {Yoshihara}}, \bibinfo
  {author} {\bibfnamefont {Y.}~\bibnamefont {Nakamura}}, \bibinfo {author}
  {\bibfnamefont {S.}~\bibnamefont {Lloyd}}, \ and\ \bibinfo {author}
  {\bibfnamefont {JS}~\bibnamefont {Tsai}},\ }\bibfield  {title} {\enquote
  {\bibinfo {title} {Quantum coherent tunable coupling of superconducting
  qubits},}\ }\href {\doibase 10.1126/science.1141324} {\bibfield  {journal}
  {\bibinfo  {journal} {Science}\ }\textbf {\bibinfo {volume} {316}},\ \bibinfo
  {pages} {723--726} (\bibinfo {year} {2007})}\BibitemShut {NoStop}%
\bibitem [{\citenamefont {Hime}\ \emph {et~al.}(2006)\citenamefont {Hime},
  \citenamefont {Reichardt}, \citenamefont {Plourde}, \citenamefont
  {Robertson}, \citenamefont {Wu}, \citenamefont {Ustinov},\ and\ \citenamefont
  {Clarke}}]{ref:hime2006solid}%
  \BibitemOpen
  \bibfield  {author} {\bibinfo {author} {\bibfnamefont {T.}~\bibnamefont
  {Hime}}, \bibinfo {author} {\bibfnamefont {P.~A.}\ \bibnamefont {Reichardt}},
  \bibinfo {author} {\bibfnamefont {B.~L.~T.}\ \bibnamefont {Plourde}},
  \bibinfo {author} {\bibfnamefont {T.~L.}\ \bibnamefont {Robertson}}, \bibinfo
  {author} {\bibfnamefont {C.-E.}\ \bibnamefont {Wu}}, \bibinfo {author}
  {\bibfnamefont {A.~V.}\ \bibnamefont {Ustinov}}, \ and\ \bibinfo {author}
  {\bibfnamefont {J.}~\bibnamefont {Clarke}},\ }\bibfield  {title} {\enquote
  {\bibinfo {title} {Solid-state qubits with current-controlled coupling},}\
  }\href {\doibase 10.1126/science.1134388} {\bibfield  {journal} {\bibinfo
  {journal} {science}\ }\textbf {\bibinfo {volume} {314}},\ \bibinfo {pages}
  {1427--1429} (\bibinfo {year} {2006})}\BibitemShut {NoStop}%
\bibitem [{\citenamefont {Cottet}\ \emph {et~al.}(2017)\citenamefont {Cottet},
  \citenamefont {Jezouin}, \citenamefont {Bretheau}, \citenamefont
  {Campagne-Ibarcq}, \citenamefont {Ficheux}, \citenamefont {Anders},
  \citenamefont {Auff{\`e}ves}, \citenamefont {Azouit}, \citenamefont
  {Rouchon},\ and\ \citenamefont {Huard}}]{ref:Cottet7561}%
  \BibitemOpen
  \bibfield  {author} {\bibinfo {author} {\bibfnamefont {N.}~\bibnamefont
  {Cottet}}, \bibinfo {author} {\bibfnamefont {S.}~\bibnamefont {Jezouin}},
  \bibinfo {author} {\bibfnamefont {L.}~\bibnamefont {Bretheau}}, \bibinfo
  {author} {\bibfnamefont {P.}~\bibnamefont {Campagne-Ibarcq}}, \bibinfo
  {author} {\bibfnamefont {Q.}~\bibnamefont {Ficheux}}, \bibinfo {author}
  {\bibfnamefont {J.}~\bibnamefont {Anders}}, \bibinfo {author} {\bibfnamefont
  {A.}~\bibnamefont {Auff{\`e}ves}}, \bibinfo {author} {\bibfnamefont
  {R.}~\bibnamefont {Azouit}}, \bibinfo {author} {\bibfnamefont
  {P.}~\bibnamefont {Rouchon}}, \ and\ \bibinfo {author} {\bibfnamefont
  {B.}~\bibnamefont {Huard}},\ }\bibfield  {title} {\enquote {\bibinfo {title}
  {Observing a quantum maxwell demon at work},}\ }\href {\doibase
  10.1073/pnas.1704827114} {\bibfield  {journal} {\bibinfo  {journal}
  {Proceedings of the National Academy of Sciences}\ }\textbf {\bibinfo
  {volume} {114}},\ \bibinfo {pages} {7561--7564} (\bibinfo {year}
  {2017})}\BibitemShut {NoStop}%
\bibitem [{\citenamefont {Gelbwaser-Klimovsky}\ \emph
  {et~al.}(2013)\citenamefont {Gelbwaser-Klimovsky}, \citenamefont {Alicki},\
  and\ \citenamefont {Kurizki}}]{ref:PhysRevE.87.012140}%
  \BibitemOpen
  \bibfield  {author} {\bibinfo {author} {\bibfnamefont {D.}~\bibnamefont
  {Gelbwaser-Klimovsky}}, \bibinfo {author} {\bibfnamefont {R.}~\bibnamefont
  {Alicki}}, \ and\ \bibinfo {author} {\bibfnamefont {G.}~\bibnamefont
  {Kurizki}},\ }\bibfield  {title} {\enquote {\bibinfo {title} {Minimal
  universal quantum heat machine},}\ }\href {\doibase
  10.1103/PhysRevE.87.012140} {\bibfield  {journal} {\bibinfo  {journal} {Phys.
  Rev. E}\ }\textbf {\bibinfo {volume} {87}},\ \bibinfo {pages} {012140}
  (\bibinfo {year} {2013})}\BibitemShut {NoStop}%
\bibitem [{\citenamefont {Myatt}\ \emph {et~al.}(2000)\citenamefont {Myatt},
  \citenamefont {King}, \citenamefont {Turchette}, \citenamefont {Sackett},
  \citenamefont {Kielpinski}, \citenamefont {Itano}, \citenamefont {Monroe},\
  and\ \citenamefont {Wineland}}]{ref:myatt2000decoherence}%
  \BibitemOpen
  \bibfield  {author} {\bibinfo {author} {\bibfnamefont {C.J.}\ \bibnamefont
  {Myatt}}, \bibinfo {author} {\bibfnamefont {B.~E.}\ \bibnamefont {King}},
  \bibinfo {author} {\bibfnamefont {Q.~A.}\ \bibnamefont {Turchette}}, \bibinfo
  {author} {\bibfnamefont {C.~A.}\ \bibnamefont {Sackett}}, \bibinfo {author}
  {\bibfnamefont {D.}~\bibnamefont {Kielpinski}}, \bibinfo {author}
  {\bibfnamefont {W.~M.}\ \bibnamefont {Itano}}, \bibinfo {author}
  {\bibfnamefont {C.}~\bibnamefont {Monroe}}, \ and\ \bibinfo {author}
  {\bibfnamefont {D.~J.}\ \bibnamefont {Wineland}},\ }\bibfield  {title}
  {\enquote {\bibinfo {title} {Decoherence of quantum superpositions through
  coupling to engineered reservoirs},}\ }\href {\doibase 10.1038/35002001}
  {\bibfield  {journal} {\bibinfo  {journal} {Nature}\ }\textbf {\bibinfo
  {volume} {403}},\ \bibinfo {pages} {269} (\bibinfo {year}
  {2000})}\BibitemShut {NoStop}%
\bibitem [{\citenamefont {Gr{\"o}blacher}\ \emph {et~al.}(2015)\citenamefont
  {Gr{\"o}blacher}, \citenamefont {Trubarov}, \citenamefont {Prigge},
  \citenamefont {Cole}, \citenamefont {Aspelmeyer},\ and\ \citenamefont
  {Eisert}}]{ref:groblacher2015observation}%
  \BibitemOpen
  \bibfield  {author} {\bibinfo {author} {\bibfnamefont {S.}~\bibnamefont
  {Gr{\"o}blacher}}, \bibinfo {author} {\bibfnamefont {A.}~\bibnamefont
  {Trubarov}}, \bibinfo {author} {\bibfnamefont {N.}~\bibnamefont {Prigge}},
  \bibinfo {author} {\bibfnamefont {G.D.}\ \bibnamefont {Cole}}, \bibinfo
  {author} {\bibfnamefont {M.}~\bibnamefont {Aspelmeyer}}, \ and\ \bibinfo
  {author} {\bibfnamefont {J.}~\bibnamefont {Eisert}},\ }\bibfield  {title}
  {\enquote {\bibinfo {title} {Observation of non-markovian micromechanical
  brownian motion},}\ }\href {\doibase 10.1038/ncomms8606} {\bibfield
  {journal} {\bibinfo  {journal} {Nature communications}\ }\textbf {\bibinfo
  {volume} {6}},\ \bibinfo {pages} {7606} (\bibinfo {year} {2015})}\BibitemShut
  {NoStop}%
\end{thebibliography}%
% Produces the bibliography via BibTeX.
%\onecolumngrid

\end{document}